\UseRawInputEncoding  

\documentclass[aps,showpacs,prd,twocolumn,nofootinbib,nobibnotes,]{revtex4-1}
\usepackage{graphicx}
\usepackage{amssymb}
\usepackage{amsmath}
\usepackage{color}
\usepackage{float}
\usepackage{ulem}
\usepackage{accents}
\usepackage{graphicx}
\usepackage{graphicx}
\usepackage{amsfonts}
\usepackage[colorlinks=true,
pdfstartview=FitV,linkcolor=blue,
citecolor=blue,urlcolor=blue,breaklinks=true]{hyperref}
\usepackage{array}
\usepackage{float}
\usepackage{placeins}
\usepackage[dvipsnames]{xcolor}
\usepackage{csquotes}
\usepackage{bbold}
\usepackage{units}
\usepackage{fixmath}

\begin{document}

\title{Symmetric and antisymmetric constitutive tensors for \\ bi-isotropic and
bi-anisotropic media}
\author{Pedro D. S. Silva}\email{pedro.dss@discente.ufma.br}\email{pdiego.10@hotmail.com}
\author{Rodolfo Casana} \email{rodolfo.casana@ufma.br}\email{rodolfo.casana@gmail.com}
\author{Manoel M. Ferreira Jr.}\email{manoel.messias@ufma.br}\email{manojr.ufma@gmail.com}
\affiliation{Departamento de F\'{\i}sica, Universidade Federal do Maranh\~{a}o, Campus Universit\'{a}rio do Bacanga, S\~{a}o Lu\'is, Maranh\~ao, 65080-805, Brazil}

\begin{abstract}

{The Maxwell equations and the constitutive relations describe the classical propagation of electromagnetic waves in continuous matter. Here, we investigate} the effects stemming from extended constitutive relations on the propagation of waves in bi-isotropic and bi-anisotropic media using a classical general {approach based on the evaluation of dispersion relations} and refractive indices. For the bi-anisotropic media, {we specify two classes of magnetoelectric parameters represented by symmetric and antisymmetric tensors.} The three cases examined have provided real and distinct refractive indices for two propagating modes, which implies birefringence. The propagating modes were also carried out in all cases. The anisotropy or birefringence effect, given by the rotatory power or phase difference, was evaluated in terms of the magnetoelectric parameters of the theory in each case. The propagation orthogonal to the vectors used to parametrize the symmetric and antisymmetric magnetoelectric tensors is described by distinct {modes, representing} a route to identify the kind of bi-anisotropic medium examined. {The group velocity and Poynting vector were also evaluated for all the cases examined to discuss the energy propagation in these anisotropic media.}

\end{abstract}

\pacs{41.20.Jb, 78.20.Ci, 78.20.Fm}
\keywords{Electromagnetic wave propagation; Optical constants;
Magneto-optical effects; Birefringence}
\maketitle

\section{\label{section1}Introduction}

As is well known, the propagation of electromagnetic waves in a { continuous and infinite medium} is described by the standard Maxwell equations in {the matter} \cite{Jackson,Zangwill},
{
\begin{subequations}
\label{maxwell-general}
	\begin{align}
	\mathbf{k}\cdot \mathbf{D}& =0\,,\quad \mathbf{k}\times \mathbf{H}+\omega
	\mathbf{D}=\mathbf{0}, \\[1ex]
	\mathbf{k}\cdot \mathbf{B}& =0\,,\quad {\bf{k}}\times {\bf{E}}- \omega \mathbf{B} ={\bf{0}} ,
	\end{align}
\end{subequations}
where we use a typical plane wave ansatz, $\mathbf{E}=\mathbf{E}%
_{0}e^{\mathrm{i}(\mathbf{k}\cdot\mathbf{r}-\omega t)}$ and $\mathbf{B}=\mathbf{B}%
_{0}e^{\mathrm{i}(\mathbf{k}\cdot\mathbf{r}-\omega t)}$.} The electric displacement and magnetic field, $\mathbf{D}$ and $\mathbf{H}$, respectively, {contain} the electromagnetic response of the matter in the form of electric polarization and magnetization, respectively. Besides, these phenomena are related to the constitutive relations involving the fields $(\mathbf{D},\mathbf{E})$ and $(\mathbf{H,B})$. {For linear, isotropic, and homogeneous dielectric matter, the constitutive relations take on the usual form,
{\begin{equation}
\mathbf{D}=\epsilon \mathbf{E}, \quad
\mathbf{H}=\mu^{-1} \mathbf{B},  \label{SCR1}
\end{equation}}
where $\epsilon$ is the electric permittivity and $\mu$ is the magnetic
permeability of the medium (constant parameters, in this case)}, given by $\epsilon =\epsilon _{0}(1+\chi^{E})$, $\mu =\mu _{0}(1+\chi ^{M})$. Here, $\chi ^{E}$ and $\chi^{M}$ represent the electric and magnetic susceptibility, respectively, contributing to the polarization, $\mathbf{P}=\epsilon _{0} \chi^{E}\mathbf{E}$, and magnetization vector, $\mathbf{M} =\chi^{M}\mathbf{H}$ \cite{Jackson,Zangwill,Landau}.  {The simplest configuration appearing in Eqs.~\eqref{SCR1} describes a medium, like water and glass, where the physical properties do not depend on the direction of the wave propagation.}

{The} complexity and diversity of electromagnetic phenomena in matter are addressed by general permittivity and permeability tensors, $\epsilon _{ij}$ and $\mu _{ij}$, written as $3 \times 3$ matrices. These tensors are suitable to describe interesting scenarios endowed with anisotropy, where the constitutive relations (\ref{SCR1}) read
\begin{equation}
D^{i}=\epsilon _{ij}E^{j}, \quad H^{i}=(\mu^{-1})_{ij}B^{j},  \label{SCR1B}
\end{equation}
with
\begin{subequations}
\label{ep-mu-0}
\begin{align}
\epsilon _{ij}& =\epsilon _{0}(\delta _{ij}+\chi _{ij}^{E}),  \label{ep0qui1}
\\
\mu _{ij}& =\mu _{0}(\delta _{ij}+\chi _{ij}^{M}),  \label{mu0qui1}
\end{align}
and $\chi _{ij}^{E}$ and $\chi _{ij}^{M}$ representing the susceptibility
tensors. The expressions in Eqs.~\eqref{ep-mu-0} include the
polarization and magnetization contributions, $P^{i}=\epsilon _{0}\chi
_{ij}^{E}E^{j}$ and $M^{i}=\chi _{ij}^{M}H^{j}$, which usually appears in the
constitutive relations as
\end{subequations}
\begin{equation}
D^{i}=\epsilon _{0}E^{i}+P^{i}, \quad
B^{i}=\mu _{0}H^{i}+\mu _{0}M^{i}.  \label{BMH1}
\end{equation}
{ For anisotropic configurations,} the tensor $\epsilon _{ij}$ describes uniaxial and biaxial crystals \cite{Landau, Bain, Fowles, Kurmanov}, which display optical activity (chirality) \cite{Condon} and birefringence \cite{Roth, Hecht}.

{Effects of anisotropy may also appear in linear electrodynamics with linear extended constitutive relations, envisaged as}
\begin{subequations}
\label{constitutive2}
\begin{align}
\mathbf{D}& =\hat{\epsilon}\, \mathbf{E}+\hat{\alpha}\,
\mathbf{B},  \label{constitutive2a} \\
\mathbf{H}& =\hat{\beta}\, \mathbf{E}+\hat{\zeta}\,\mathbf{B },
\label{constitutive2b}
\end{align}
\end{subequations}
where $\hat{\epsilon}=[\epsilon_{ij}]$, $\hat{\alpha}=[\alpha_{ij}]$, $\hat{\beta}=[\beta_{ij}]$, and $\hat{\zeta}=[\zeta_{ij}]$
{represent, in principle, $3\times 3$ complex matrices.} Such expressions above contain additional magnetoelectric responses of the medium to electromagnetic fields: $\hat{\alpha}$ measures the electric response to the
magnetic field and $\hat{\beta}$ represents the magnetic response to the electric field. In this generalized context, if the relations (\ref{BMH1}) remain valid, the polarization and magnetization vectors receive contributions from the magnetic and electric sectors, respectively, being given as
\begin{subequations}
	\label{polarization-magnetization-1}
	\begin{align}
	P^{i}& =\epsilon _{0}\chi _{ij}^{E}E^{j}+\alpha _{ij}B^{j},  \label{p1} \\
	M^{i}& =\chi _{ij}^{M}H^{j}+\tilde{\beta}_{ij}E^{j},  \label{p2}
	\end{align}
	\end{subequations}		
where it holds that $\hat\zeta=\hat\mu^{-1},\quad \hat\beta =-\mu _{0}\hat\mu^{-1} \tilde{\beta}$.

In order to ensure the energy conservation of the electromagnetic field in a medium where the constitutive relations (\ref{constitutive2a}) and (\ref{constitutive2b}) prevail, the Poynting theorem leads to the following set of relations for the complex matrices:
\begin{subequations}
	\label{relation-parameter0}
	\begin{align}
	\hat{\epsilon} &= \hat{\epsilon}^{\dagger},  \label{relation-parameter1}\\
	 \hat{\mu}^{-1}&=( \hat{\mu}^{-1})^{\dagger},  \label{relation-parameter2}\\
	\hat{\alpha}&=-\hat{\beta}^{\dagger} \label{relation-parameter3} .
	\end{align}
\end{subequations}
{For details, see Sec.~\ref{section-extended-constitutive}} and Refs. \cite{Sihvola,Kong, Kamenetskii}. Relation (\ref{relation-parameter3}) will be crucially relevant in the analysis of the present work, as will be clear in the next sections.

The simplest version of the relations in Eqs.~\eqref{constitutive2},
\begin{equation}
\begin{pmatrix}
\mathbf{D} \\
\mathbf{H}	\end{pmatrix}	=	\begin{pmatrix}
\epsilon  & \alpha  \\
\beta  & {\zeta }	\end{pmatrix}	\begin{pmatrix}
\mathbf{E} \\
\mathbf{B}	\end{pmatrix}	\label{Matrix2}
\,,
\end{equation}
{includes} $\epsilon $, $\alpha $, $\beta$, and $\zeta$ as single parameters and
describes the physics of bi-isotropic media (the most general linear,
homogenous and isotropic materials \cite{ Sihvola, Kong}), {corresponding to the case in which the matrices $\hat{\alpha}=[\alpha_{ij}]$, $\hat{\beta}=[\beta_{ij}]$ are diagonal and isotropic.}
In this case,  in order to {be consistent with} energy conservation, the relations \eqref{Matrix2} yield
\begin{equation}
\alpha=-\beta^{*}.  \label{SCR2}
\end{equation}
In the configuration the constitutive relations (\ref{Matrix2}) have the form $\mathbf{D} ={\epsilon}\mathbf{E}+{\alpha}\mathbf{H}$ and $\mathbf{B} ={\beta}\mathbf{E}+{\zeta}\mathbf{H}$,
the condition (\ref{SCR2}) becomes $\alpha=\beta^{*}=\psi +i \kappa$,  where $\psi$ is the Tellegen coefficient and $\kappa$ is
the chirality coefficient \cite{Aladadi}. See Eqs.~\eqref{poynting-23} for details.
The bi-isotropic relations (\ref{Matrix2}) have been much studied {in both theoretical \cite{Sihvola1,Sihvola2,Sihvola3, Nieves, Gauthier} and applied {aspects} \cite{Rado, Aladadi, Jelinek},} being also important to address {optical properties \cite{Chang, Zou} and other properties of
topological insulators \cite{Urrutia, Urrutia2, Lakhtakia, Winder,
Li,Li1,Tokura}. Bi-isotropic relations are relevant} for axion electrodynamics \cite{citekey, Sekine, Tobar, BorgesAxion}, construction of optical isolators from chiral materials \cite{Silveirinha}, the Casimir effect in chiral media \cite{Casimir}, and other applications \cite{Darinskii}. Furthermore, bi-anisotropic ``chiral materials", described by relations (\ref{constitutive2}) involving anisotropic tensors, were employed to
investigate relativistic electron gas \cite{Carvalho}, time-dependent magnetoelectric parameters \cite{Lin}, Weyl semimetals \cite{Halterman, Zu}, magnetized materials \cite{ Krupka1, Krupka2}, and anisotropic
dispersion relations \cite{Hillion, Yakov, Damaskos}. {It is also worthy to mention some effects engendered by the anisotropic magnetoelectric parameters, corresponding to the off-diagonal elements of the matrices  $\hat{\alpha}=[\alpha_{ij}]$ and $\hat{\beta}=[\beta_{ij}]$. Nondiagonal terms, for instance, were examined in the discovery of electromagnons in perovskites, which revealed an absorption difference of light propagating in opposite directions (directional dichroism) \cite{Takahashi}. Magnetoelectric diagonal (and anisotropic) coefficients were investigated in the context of multiferroic materials, where they induced a light polarization rotation angle \cite{Kurumaji}.}

In extended scenarios, the constitutive tensors of relations (\ref{constitutive2}) may also depend on the space coordinates, standing for the description of {nonhomogeneous} bi-isotropic and/or bi-anisotropic media \cite{Novitsky}. These tensors can present dependence on the magnitude of the electromagnetic fields as well, $\hat{\epsilon}=\hat{\epsilon}({E, B})$ and  $\hat{\mu}=\hat{\mu}(E, B)$, a kind of approach which accounts for birefringence in nonlinear electrodynamics \cite{Lorenci1}, allowing to recover the Kerr and Cotton-Mouton effects in particular configurations \cite{Lorenci2}. A more involved and general nonlinear construction, where the magnetoelectric coefficients exhibit dependence at second order on the electromagnetic field components, was recently examined  \cite{Lorenci3}.

Generalized constitutive relations can also be envisaged for the current density as an extension of the standard Ohm's law. Such relations can be written as $J^{i} = \sigma E^{i}+ \sigma^{B}_{ij} B^{j}$, where $\sigma$ is the usual Ohmic conductivity and $\sigma^{B}_{ij}$ is a general magnetic conductivity tensor. An isotropic tensor, $\sigma^{B}_{ij}=\Sigma \delta_{ij}$, stands for the chiral magnetic effect (CME) \cite{Kharzeev1, Qiu, Fukushima}. Isotropic and anisotropic symmetric and antisymmetric conductivity tensors were examined in Ref. \cite{Pedro1}. The antisymmetric parametrization of $\sigma^{B}_{ij}$ also has found realization in some Weyl semimetals \cite{Kaushik}.

Another possible extension occurs in the context of a Lorentz-violating
anisotropic electrodynamics~\cite{Tobar1, Bailey}, with constitutive
relations written as
\begin{equation}
\begin{pmatrix}
\mathbf{D} \\
\mathbf{H}%
\end{pmatrix}=
\begin{pmatrix}
\epsilon \mathbb{1}+ { \hat{\kappa} _{DE}} &   {\hat{\kappa} _{DB} } \\
&  \\
{ \hat{\kappa} _{HE} } & {\mu }^{-1}\mathbb{1}+  {\hat{\kappa} _{HB}  }%
\end{pmatrix}%
\begin{pmatrix}
\mathbf{E} \\
\mathbf{B}%
\end{pmatrix}%
\,,  \label{eq15}
\end{equation}%
{where $\hat{\kappa}_{DE}$, $\hat{\kappa}_{DB}$, $\hat{\kappa}_{HE}$, and $\hat{\kappa}_{HB}$} are
dimensionless $3\times3$ matrices composed of vacuum, {$ \hat{\kappa} _{DE}^{vac}$, $%
\hat{\kappa} _{DB}^{vac}$, $\hat{\kappa} _{HE}^{vac}$, and $\hat{\kappa} _{HB}^{vac},$} and matter
pieces, $\hat{\kappa} _{DE}^{matter}$, $\hat{\kappa} _{DB}^{matter}$, $\hat{\kappa}
_{HE}^{matter}$, and $\hat{\kappa}_{HB}^{matter}$. These generalized scenarios lead
to unusual electrodynamics where magnetoelectric parameters stemming from Lorentz symmetry violation appear {in matter or vacuum,} giving rise to interesting effects potentially related to the phenomenology of new materials. A classical field theory approach to the description of wave propagation in a continuous chiral medium supporting higher-order derivative Lorentz-violating electrodynamics was recently examined \cite{Pedro2}.

In this work, we investigate the possible effects stemming from extended linear constitutive relations (\ref{constitutive2a}) and (\ref{constitutive2b}), assuming isotropic electric permittivity and magnetic permeability,
$\zeta _{ij}={\mu}^{-1}\delta_{ij}$, $\epsilon_{ij}=\epsilon\delta_{ij}$, and that the tensors $\alpha _{ij}$ and $\beta _{ij}$ may be described by symmetric and antisymmetric parametrizations.  {For the three cases investigated, we have obtained general dispersive equations which provide the refractive indices for any propagation direction. For the anisotropic magnetoelectric tensors, we have worked out specific solutions for the particular propagation axis in order to discuss the optical repercussions.} {More specifically, the bi-anisotropic symmetric constitutive relations are parametrized in terms of a 3-vector ${\bf{d}}$,
	\begin{align}
		{\bf{D}} = \epsilon {\bf{E}} +\tilde{\alpha} {\bf{d}} ({\bf{d}}\cdot {\bf{B}}), \quad {\bf{H}} = \mu^{-1} {\bf{B}} + \tilde{\beta} {\bf{d}} ({\bf{d}} \cdot {\bf{E}}),
	\end{align}
for which we discuss the dispersion relations, refractive indices, and birefringence for special configurations where the propagation vector direction is along and perpendicular to the vector ${\bf{d}}$. The antisymmetric constitutive relations are parametrized in terms of two 3-vectors ${\bf{a}}$ and ${\bf{b}}$, 
	\begin{align}
		{\bf{D}} = \epsilon {\bf{E}} + {\bf{a}} \times {\bf{B}}, \quad {\bf{H}} = \mu^{-1} {\bf{B}} + {\bf{b}} \times {\bf{E}},
	\end{align}
{satisfying ${\bf{b}}={\bf{a}}^{*}$ and} used to describe the particular scenarios where the propagation direction is longitudinal and orthogonal to the vector ${\bf{a}}$.}

The paper is outlined as follows: In Sec. \ref
{section-extended-constitutive}, we present the basic formalism for
obtaining the dispersion relations and refractive indices in a general
scenario of extended constitutive relations. {In Sec.~\ref{bi-isotropic-case},} we discuss the electromagnetic wave propagation in the bi-isotropic case. In the sequel, we focus on the isotropic-anisotropic constitutive relations, examining symmetric {(see Sec.~\ref{bianisotropic_Symm})} and antisymmetric {(see Sec.~\ref{bianisotropic_Asymm}) }configurations for the tensors $\alpha_{ka}$ and $\beta_{ka}$. Finally, in Sec.~\ref{final-remarks}, we summarize our results. {Throughout the paper, we use natural units}.

{
\section{\label{section-extended-constitutive} Dispersion relations for
bi-isotropic and bi-anisotropic media described by extended linear constitutive relations}}

In this section, we start from the Maxwell equations in a homogeneous ponderable {nonconducting} medium endowed with general linear constitutive relations in order to obtain the dispersion relations, which provide the refractive index and the propagating modes. {From Eq. (\ref{maxwell-general}), Amp\`{e}re's law} {reads}
\begin{equation}
{\varepsilon_{ijk}} k^{j}H^{k}+\omega D^{i}=0,  \label{ex1}
\end{equation}%
{where $\varepsilon_{ijk}$ is the tridimensional Levi-Civita symbol.} Replacing the constitutive relations (\ref{constitutive2a}) and (\ref%
{constitutive2b}) in Eq.~\eqref{ex1}, one has
\begin{equation}
{\varepsilon_{ijk}} k^{j}\left( \beta_{ka}E^{a}+\zeta _{ka}B^{a}\right) +\omega
\epsilon_{ij}E^{j}+\omega \alpha _{ij}B^{j}=0,
\end{equation}%
Employing now Faraday's law, $\omega \mathbf{B}=\mathbf{k\times E}$, {one obtains an equation totally in terms of the electric field, 
\begin{eqnarray}
0&=&\varepsilon_{ijk}\varepsilon_{amn} \zeta_{ka}k^{j}k^{m} E^{n} +\omega^{2} \epsilon_{ij}E^{j} + \nonumber \\
&&+ \omega \varepsilon_{jmn}\alpha_{ij}k^{m}E^{n}+ \omega \varepsilon_{ijk} \beta_{ka}k^{j}E^{a} .
\label{ex2}
\end{eqnarray}   }
{Let us consider that} the medium has isotropic both the electric permittivity and magnetic permeability,
\begin{equation}
\zeta _{ka}={\mu }^{-1}\delta _{ka},\text{ \ \ }\epsilon _{ij}=\epsilon
\delta _{ij},  \label{ex3}
\end{equation}%
in such a way that the anisotropy, typical of ``chiral'' media, is allowed to exist in
the magnetoelectric coefficients. Hence, Eq.\eqref{ex2} becomes
\begin{equation}
\left[ \mathbf{k}\times \left(\mathbf{k}\times \mathbf{E}\right)\right]
^{i}+\omega ^{2}\mu \bar{\epsilon}_{ij}E^{j}=0,  \label{ex8}
\end{equation}%
where
\begin{equation}
\bar{\epsilon}_{in}(\omega )=\epsilon\delta _{in}-\frac{1}{\omega }  \left( \beta
_{kn}   {\varepsilon_{imk}}  +\alpha _{ij}   {\varepsilon_{jmn}}  \right) k_{m},  \label{ex7}
\end{equation}%
defines the frequency-dependent extended permittivity tensor, which carries
the electric and magnetic response of the medium. Equation \eqref{ex8} is also {cast in the form}
\begin{equation}
\left[ k^{2}\delta _{ij}-k_{i}k_{j}-\omega ^{2}\mu \bar{\epsilon}_{ij}\right]
E^{j}=0.  \label{ex9}
\end{equation}%
For a general anisotropic continuous scenario, we write $\mathbf{k}=\omega
\mathbf{n}$ where $\mathbf{n}$ is a vector pointing along the direction of
the wave vector and yields the refractive index: $n=+\sqrt{\mathbf{n}^{2}}$.
Here we consider that the index $n$ is nonnegative and $\sqrt{\mathbf{n}^{2}}
$ instead of $|\mathbf{n}|$, in order to permit complex refractive indices.
The refractive indices with negative real parts, related to metamaterials,
are not considered here. Hence, Eq.\eqref{ex9} becomes
\begin{equation}
M_{ij}E^{j}=0,  \label{ex11}
\end{equation}
where the tensor $M_{ij}$ reads
\begin{equation}
M_{ij}=n^{2}{\delta }_{ij}-n_{i}n_{j}-\mu \bar{{\epsilon }}_{ij},
\label{ex12}
\end{equation}%
and $\bar{{\epsilon }}_{ij}$ is given by Eq.\eqref{ex7}. This set of equations
has a nontrivial solution for the electric field if the determinant of the
matrix $M_{ij}$ vanishes. Such a condition provides the dispersion relations
that govern the wave propagation in the medium. In the case of standard
media described by anisotropic tensors $\epsilon_{ij}$ and $\zeta_{ij}$, and
with no extensions on the constitutive relations, $\alpha_{ij}=0$ and $%
\beta_{ij}=0$, the dispersion relation can be found in Refs.~\cite{Yakov, Damaskos}.


{We next examine the propagation} of electromagnetic waves in a
dielectric medium under the validity of anisotropic extended dispersion
relations of the form
\begin{subequations}
\label{constitutive-relations-general-2}
\begin{eqnarray}
D^{i} &=&\epsilon \delta_{ij}E^{j}+\alpha _{ij}B^{j},  \label{DEB4a} \\
H^{i} &=&\beta _{ij}E^{j}+\mu^{-1}\delta_{ij}B^{j},  \label{HEB4a}
\end{eqnarray}
{where Eqs.~(\ref{ex3}) were considered.} These relations may be considered isotropic-anisotropic since they contain isotropic electric permittivity and magnetic permeability, but anisotropic magnetoelectric tensors, $\alpha_{ij},\beta _{ij}$.

{First, we present the general conditions on the constitutive tensors in order to ensure energy conservation in the system. The Poynting theorem is given by \cite{Jackson}
\begin{align}
\nabla \cdot {\bf{S}} &= -\frac{\mathrm{i}\omega}{2} \left({\bf{E}} \cdot {\bf{D}}^{*} - {\bf{H}}^{*} \cdot {\bf{B}}\right) - \frac{ ({\bf{J}}^{*} \cdot {\bf{E}})}{2}, \label{eq:poynting-theorem-1}
\end{align}
where
\begin{align}
{\bf{S}}=\frac{1}{2} \left({\bf{E}}\times {\bf{H}}^{*} \right), \label{eq:poynting-theorem-2}
\end{align}
is the Poynting vector. The real part of Eq.~\eqref{eq:poynting-theorem-1} yields the energy conservation law for the system. In the absence of sources and considering there is no flux of energy density, the energy conservation condition is
\begin{align}
\mathrm{Re} \left[ \mathrm{i}\omega \left( {\bf{D}}^{\dagger} {\bf{E}} - {\bf{H}}^{\dagger} {\bf{B}} \right) \right] =0 . \label{eq:poynting-theorem-3}
\end{align}
Depending on the form of the constitutive relations, constraints on the parameters describing the medium (compatible with energy conservation) are obtained. Indeed, replacing the constitutive relations in Eqs.~\eqref{constitutive-relations-general-2} into Eq.~\eqref{eq:poynting-theorem-3} yields
\begin{align}
0 &=\mathbf{E}^{\dag }\left( \hat{\epsilon}^{\dag }-\hat{\epsilon}\right)
\mathbf{E}-\mathbf{B{^{\dag }}}\left( \hat{\zeta}\mathbf{{^{\dag }}}-\hat{\zeta}\right) \mathbf{B}  \notag \\
&+\mathbf{B}^{\dag }\left( \hat{\beta}+\hat{\alpha}^{\dag }\right) \mathbf{E}-\mathbf{E{^{\dag }}}\left( \hat{\beta}\mathbf{{^{\dag }}}+\hat{\alpha}\right) \mathbf{B}, \label{eq:poynting-theorem-4}
\end{align}
which establishes a general relation involving all constitutive tensors with the electromagnetic fields. A simple route to ensure energy conservation is to set
\begin{align}
\hat{\epsilon}^{\dagger}= \hat{\epsilon} ,\quad \hat{\zeta}^{\dagger}=\zeta, \quad \hat{\beta}=-\hat{\alpha}^{\dagger} . \label{eq:poynting-theorem-5}
\end{align}
The last condition will be relevant in the discussions of the next sections, as we will see. In the following, we write the dispersion relations} from which we obtain the refractive indices and the propagating modes for some special configurations of $\alpha_{ij},\beta _{ij}$.

{In the next sections, we study the dispersion relations, refractive indices, propagating modes, group velocity, phase velocity, and Poynting vector of the electromagnetic waves for the bi-isotropic and bi-anisotropic linear media. }

{\section{Bi-isotropic case \label{bi-isotropic-case}}}

In the context of the constitutive relations
\eqref{constitutive-relations-general-2}, we begin considering the total
symmetric isotropic configuration, where the quantities $\hat{\alpha}$
and $\hat{\beta}$ are given by
\end{subequations}
\begin{equation}
\alpha _{ij} =\alpha \delta _{ij}, \quad \beta _{ij} =\beta \delta _{ij} ,
\label{isot5}
\end{equation}
with $\alpha$, $\beta \in \mathbb{C}$. The condition $\alpha_{ij}=-\beta_{ij}^{\dagger}$, when applied on parametrization \eqref{isot5}, yields
\begin{align}
\beta^{*}=-\alpha. \label{isotropic-case-1-1}
\end{align}
In this case, the constitutive relations take on the typical bi-isotropic form,
\begin{subequations}
\label{constitutive-relations-biisotropic-1}
\begin{align}
\mathbf{D}& =\epsilon \mathbf{E}+\alpha \mathbf{B},
\label{constitutive-iso1D} \\
\mathbf{H}& =\frac{1}{\mu }\mathbf{B}+\beta \mathbf{E},
\label{constitutive-iso2D}
\end{align}%
which represent the simplest linear connection between $(\mathbf{D},\mathbf{H%
})$ and $(\mathbf{E},\mathbf{B})$. As already mentioned, such relations play
a relevant role in topological insulators \cite{Chang, Urrutia, Urrutia2,
Lakhtakia, Winder, Li,Li1} and axion systems \cite{Sekine, Tobar,
BorgesAxion}.

Inserting Eq. (\ref{isot5}) in Eq.~\eqref{ex7}, one obtains
\end{subequations}
\begin{equation}
\bar{\epsilon}_{ij}= \epsilon \delta _{ij}+(\alpha +\beta ) { \varepsilon_{ijm}}  n_{m},
\label{iso4}
\end{equation}%
where the last term on the right-hand side represents the ``magnetic-electric" response of
the medium. As we have started with isotropic tensors, $\epsilon \delta
_{ij} $, $\mu ^{-1}\delta _{ij}$, $\alpha \delta _{ij}$, and $\beta \delta _{ij}$%
, any effective arising anisotropy comes from the extended structure of the
constitutive relations \eqref{constitutive-relations-biisotropic-1}. In
this case, the tensor $M_{ij}$ [Eq.~\eqref{ex12}] has the form
\begin{align}
{M  \equiv [M_{ij}] } &= \mathcal{N}  - \mu (\alpha+\beta) \begin{pmatrix}
0 & n_{3} & - n_{2} \\
- n_{3} & 0 & n_{1} \\
n_{2} & - n_{1} & 0
\end{pmatrix} ,  \label{eq62}
\end{align}
where
\begin{equation}
\mathcal{N}=\begin{pmatrix}
n_{2}^{2}+n_{3}^{2}-\mu\epsilon  & -n_{1}n_{2} & -n_{1}n_{3} \\
-n_{1}n_{2} & n_{1}^{2}+n_{3}^{2}-\mu\epsilon  & -n_{2}n_{3} \\
-n_{1}n_{3} & -n_{2}n_{3} & n_{1}^{2}+n_{2}^{2} -\mu\epsilon
\end{pmatrix} .\label{Nij1}
\end{equation}	

Requiring $\mathrm{det}[M_{ij}]=0$, one gets
\begin{equation}
n^{4}-n^{2}\left[ 2\mu\epsilon  -\mu ^{2}(\alpha +\beta )^{2}\right] +\mu ^{2}\epsilon^{2}=0.  \label{iso6}
\end{equation}%
Solving for $n^{2}$, we obtain the following refractive indices
\begin{align}
n_{\pm }^{2}=& \mu{\epsilon}- 2Z \pm \mathrm{i}\mu (\alpha +\beta )%
\sqrt{\mu\epsilon -Z},  \label{iso7}
\end{align}
where
\begin{equation}
Z=\frac{\mu ^{2}(\alpha +\beta )^{2}}{4}.  \label{iso-7-1}
\end{equation}
Thereby the corresponding $n_{\pm }$ {read}
\begin{equation}
n_{\pm }=\sqrt{\mu\epsilon -Z} \pm \mathrm{i}\sqrt{Z},
\label{iso10}
\end{equation}
where we have considered only the indices with a positive real piece in order
to avoid metamaterial behavior. {The refractive indices \eqref{iso10} are valid (and are equal) for any propagation direction since the bi-isotropic case does not have a preferred direction that could be represented by a constant vector. In spite of that, the system may manifest {anisotropic} effect (circular {birefringence}) due to the way the fields are coupled. Such an effect will be examined ahead.}

 The refractive indices can also be obtained
by diagonalizing the electric permittivity and setting each eigenvalue equal
to $n^{2}/\mu $. The eigenvalues $\epsilon _{a}$ ($a=1,2,3$) fulfill $\bar{%
\epsilon}_{ij}e_{a}^{j}=\epsilon _{a}e_{a}^{i}$, where $\mathbf{e}_{a}$
represent the eigenvectors. Diagonalizing the matrix of the operator ${\bar{%
\epsilon}}$ [Eq.~\eqref{iso4}], one finds the following eigenvalues:
\begin{align}
\epsilon _{1}& =\epsilon ,  \label{iso11}
\\
\epsilon _{2,3}& \equiv \epsilon _{\pm }=\epsilon \pm \mathrm{i}(\alpha +\beta )n,  \label{iso12}
\end{align}%
associated with the eigenvectors
\begin{align}
{\mathbf{e}}_{1}& =\frac{\mathbf{n}}{n} ,  \label{iso16} \\
\mathbf{e}_{2(3)}& = \frac{1}{n\sqrt{2(n^{2}_{1}+n_{3}^{2})}}
\begin{pmatrix}
n_{3}n \pm \mathrm{i}n_{1}n_{2} \\
\mp \mathrm{i} (n_{1}^{2}+n^{2}_{3} ) \\
\pm \mathrm{i} n_{2}n_{3} - n_{1}n%
\end{pmatrix}
.  \label{iso17}
\end{align}
Eigenvalues (\ref{iso11}) and (\ref{iso12}) are associated with the
refractive indices
\begin{align}
n^{2}& =\mu  \epsilon ,\label{iso13} \\
n_{\pm }^{2}& =\mu \epsilon \pm \mathrm{i}\mu (\alpha +\beta )n.  \label{iso14}
\end{align}%
We note that Eq.\eqref{iso13} represents the refractive index of an {isotropic
dielectric medium}. On the other hand, Eq.\eqref{iso14} recovers the
result of Eq.~\eqref{iso10}, meaning that only the eigenvalues $\epsilon _{2}$
and $\epsilon _{3}$ correspond to the refractive indices of the medium $%
n_{+} $ and $n_{-}$, respectively. This approach of finding the refractive
indices $n$ via the relation $n^{2}=\mu \epsilon _{a}(n)$, where $\epsilon
_{a}$ stands for the eigenvalues of the electric permittivity ${\bar{\epsilon%
}}_{ij}$, only works when the electric field is orthogonal to the
propagation direction. Here, such a condition is guaranteed by the Gauss' law ${\bf{k}}\cdot {\bf{D}} =0$, where the electric displacement vector, given by
\begin{align}
{\bf{D}} &= \epsilon {\bf{E}} + \frac{\alpha}{\omega} {\bf{k}}\times {\bf{E}},\label{DDiso}
\end{align}
provides ${\bf{k}}\cdot {\bf{E}}=0$. Then for a general vector $\mathbf{n}=%
\mathbf{k}/\omega$, the related propagating electric field, ${\mathbf{E}}%
_{a} $, satisfies ${\mathbf{n}}\cdot {\mathbf{E}}_{a}=0$. %
This way,  Eqs.\eqref{ex11} and \eqref{ex12} simplify to
\begin{equation}
\left[ n^{2}\delta _{ij}-\mu \bar{\epsilon}_{ij}\right] E^{j}=0,
\end{equation}
or $n^{2}\delta _{ij}=\mu \bar{\epsilon}_{ij}$, creating the straightforward
correspondence between $n^{2}$ and $\bar{\epsilon}_{ij}$ eigenvalues, that
is, $n^{2}=\mu \epsilon _{a}(n)$. This is the reason by which the
eigenvectors ($e_{a})$ represent the electric field modes, $\mathbf{E}%
_{a}\sim \mathbf{e}_{a}$. Note that it also holds that ${\mathbf{n}}\cdot {%
\mathbf{e}}_{a}=0$. This situation is clearly illustrated in the present
case. In fact, the three normalized eigenvectors given in Eqs. \eqref{iso16} and %
\eqref{iso17} are linearly independent {and obey}
\begin{equation}
\mathbf{e}_{1}\cdot \mathbf{e}_{2}^{\ast }=\mathbf{e}_{1}\cdot \mathbf{e}%
_{3}^{\ast }=\mathbf{e}_{2}\cdot \mathbf{e}_{3}^{\ast }=0.
\end{equation}
In particular, ${\mathbf{e}}_{2}$ and ${\mathbf{e}}_{3}$ are orthogonal to ${%
\mathbf{e}}_{1}\sim {\mathbf{n}}$, thus indicating the transversality of the
propagating modes, $\mathbf{E}_{2}\sim \mathbf{e}_{2,}$ $\mathbf{E}_{3}\sim
\mathbf{e}_{3}$, whose eigenvalues yield the correct refractive indices $%
n_{\pm }$ [see Eq.\eqref{iso10}]. On the other hand, ${\mathbf{n}}\cdot {%
\mathbf{e}}_{1}$ is nonzero, meaning that the eigenvalue $\epsilon _{1}$
does not yields a physical refractive index.

\bigskip

\subsection{Propagation modes}

As already explained, the electric field of the
propagating modes is given by solution (\ref{iso17}). So, let us choose a
a convenient coordinate system where
\begin{equation}
\mathbf{n}=(0,0,n_{3}), \label{iso-prop-1}
\end{equation}%
with which the eigenvectors (\ref{iso17}) {are}
\begin{equation}
\mathbf{e}_{2(3)}=\frac{1}{\sqrt{2}}%
\begin{pmatrix}
1 \\
\mp \mathrm{i} \\
0%
\end{pmatrix}%
,  \label{iso-prop-3}
\end{equation}%
where $-\mathrm{i}$ represents a right-handed circular polarization (RCP)
and $+\mathrm{i}$ a left-handed circular polarization (LCP), respectively.
We can easily show that the same result stems directly from \eqref{ex11}. By
replacing the simple choice of Eq.(\ref{iso-prop-1}) in the matrix (\ref{eq62}),
\begin{equation}
{M}=%
\begin{pmatrix}
n_{3}^{2}-\mu \epsilon &  & -\mu (\alpha +\beta )n_{3} &  & 0 \\
\mu (\alpha +\beta )n_{3} &  & n_{3}^{2}-\mu \epsilon &  & 0 \\
0 &  & 0 &  & -\mu \epsilon%
\end{pmatrix}%
,  \label{matrix-isotropic-modes-1}
\end{equation}%
and implementing the refractive indices (\ref{iso7}), the condition $%
M_{ij}E^{j}=0$ provides the following normalized solutions of the electric
field of the propagating modes:
\begin{equation}
\hat{\mathbf{E}}_{\pm }=\frac{1}{\sqrt{2}}%
\begin{pmatrix}
1 \\
\pm \mathrm{i} \\
0%
\end{pmatrix}%
,  \label{iso-prop-2}
\end{equation}%
where
$\hat{\bf{E}}_{+}$ and $\hat{\bf{E}}_{-}$ represent
the LCP and RCP vectors, respectively. Solution (\ref{iso-prop-2}) does
not depend on the nature (real or complex) of the parameters $\alpha$ and
$\beta$, in such a way it will be valid for all the cases examined in this
section. The equality between solution (\ref{iso-prop-2}) and %
Eq.~\eqref{iso-prop-3}, $\mathbf{e}_{3(2)}\equiv \hat{\mathbf{E}}_{\pm }$,
confirms the approach here developed.

We point out that the circular polarization solution (\ref{iso-prop-2}) is not exclusive of the $z$-propagation direction. Indeed, taking on the propagation in the $x$-axis, $\mathbf{n}=(n_{1},0,0)$, matrix (\ref{eq62}) takes the form,
	\begin{equation}
	{M}=%
	\begin{pmatrix}
	-\mu \epsilon &  & 0  &  & 0 \\
	0 &  & n_{1}^{2}-\mu \epsilon &  & -\mu (\alpha +\beta )n_{1} \\
	0 &  & \mu (\alpha +\beta )n_{1} &  & n_{1}^{2}-\mu \epsilon%
	\end{pmatrix},  \label{matrix-isotropic-modes-1B}
	\end{equation}
whose associated modes,
\begin{equation}
\hat{\mathbf{E}}_{\pm }=\frac{1}{\sqrt{2}}%
\begin{pmatrix}
0 \\
\pm \mathrm{i} \\
1%
\end{pmatrix}%
,  \label{iso-prop-2B}
\end{equation}
also correspond to transversal circularly polarized waves.

\subsection{Optical effects of complex magnetoelectric parameters in dielectrics}

Since we have already found the refractive indices and the polarization of the propagating modes, it is necessary to examine the physical behavior brought about by the constitutive relations (\ref{constitutive-iso1D}) and (\ref{constitutive-iso2D}) on a conventional dielectric substrate. In the
limit $(\alpha +\beta )\rightarrow 0$, one recovers the refractive index
of an isotropic {dielectric medium, given by Eq. \eqref{iso13},}
\begin{equation}
n_{\pm }^{2}=\mu \epsilon .  \label{iso8}
\end{equation}
Equation \eqref{iso10} provides
\begin{equation}
n_{\pm }=\sqrt{\mu \epsilon -\frac{\mu ^{2}(\alpha +\beta )^{2}}{4}}\pm
\mathrm{i}\frac{\mu (\alpha +\beta )}{2}.  \label{iso19}
\end{equation}

Now we examine the refractive indices  (\ref{iso19}) in two
cases: (a) $\alpha$, $\beta \in \mathbb{C}$ and (b) $\alpha$, $\beta \in
\mathbb{R}$.
For $\alpha$ and $\beta$ complex, one can write
\begin{equation}
\alpha=\alpha' +\mathrm{i} \alpha'', \quad \beta=\beta' +\mathrm{i} \beta'',
\end{equation}
where $\alpha'=\mathrm{Re}[\alpha]$, $\alpha''=\mathrm{Im}[\alpha]$, $\beta'=\mathrm{Re}[\beta]$ and $\beta''=\mathrm{Im}[\beta]$. Condition (\ref{isotropic-case-1-1}) implies
\begin{equation}
\alpha'=-\beta', \quad \alpha''=\beta'',
\end{equation}
so that  $\alpha+\beta= 2{\mathrm{i}\alpha''}$. Therefore, Eq.~\eqref{iso19} is rewritten as
\begin{equation}
n_{\pm }=\sqrt{\mu \epsilon +\mu ^{2}{\alpha ^{\prime \prime }}^{2}}\mp \mu
\alpha ^{\prime \prime },  \label{isotropic-case-1-2}
\end{equation}%
which are real, positive, and cause birefringence. Since the
polarization modes are circularly polarized vectors [see Eq.~\eqref{iso-prop-2}],
the birefringence effect can be evaluated in terms of the rotatory power
(see Appendix \ref{AppendixA}), defined as
\begin{equation}
\delta =-\frac{[\mathrm{Re}(n_{+})-\mathrm{Re}(n_{-})]\omega }{2}\
\label{eq:rotatory-power1A}
\end{equation}%
Hence, using indices (\ref{isotropic-case-1-2}), the rotatory power is
\begin{equation}
\delta =\mu \omega \alpha ^{\prime \prime }.  \label{isotropic-case-1-3}
\end{equation}

Such a birefringence effect [Eq.~(\ref{isotropic-case-1-3})] is a consequence of $%
(\alpha+\beta)=2\mathrm{i}{\alpha}^{\prime\prime}$. Therefore, it only occurs when the constitutive
parameters possess an imaginary piece. On the other hand, for $\alpha$, $\beta \in \mathbb{R}$, one has simply $\beta=-\alpha$, $\alpha''=0$, and no birefringence takes place.
This is the case of the topological insulators bi-isotropic scenario \cite{Chang,
Urrutia, Urrutia2, Lakhtakia, Winder, Li,Li1}, whose constitutive
relations are
\begin{subequations}
\label{iso20A}
\begin{eqnarray}
\mathbf{D} &=&\epsilon \mathbf{E}-\alpha _{0}\mathbf{B},  \label{iso20} \\
\mathbf{H} &=&\frac{\mathbf{B}}{\mu }+\alpha _{0}\mathbf{E},  \label{iso21}
\end{eqnarray}
\end{subequations}
with $\alpha _{0}=e^{2}/4\pi \hbar$ and $e$ being the elementary electric
charge.
For Eq.\eqref{iso20A}, one has  $(\alpha+\beta)=0$, so that no birefringence is provided.

Concerning topological insulators, quantum effects of bulk interband excitations may generate strong Faraday rotation associated with a type of optical activity described by the Verdet constant \cite{Ohnoutek, Liang}. The quantum origin of this effect does not represent a contradiction with the classical absence of birefringence for topological insulators here remarked.

\subsection{Group velocity, phase velocity, and Poynting vector}

{Using ${\bf{n}}={\bf{k}}/\omega$ in Eq. (\ref{iso10}), one finds
\begin{align}
\omega_{\pm} &= \frac{k}{\sqrt{\mu\epsilon - Z} \pm \mathrm{i} \sqrt{Z}}. \label{eq:group-velocity-isotropic-case-1}
\end{align}
To assess the group and phase velocities, we need to consider the nature (real or complex) of the magnetoelectric parameters.
\begin{itemize}
\item For $\alpha$, $\beta \in \mathbb{C}$, it holds that $(\alpha+\beta) = 2\mathrm{i} \alpha''$ and $Z=-\mu^{2}\alpha''^{2}$, so that
\begin{align}
\omega_{\pm} &= \frac{k}{\sqrt{\mu\epsilon + \mu \alpha''^{2}} \mp \mu\alpha''}. \label{eq:group-velocity-isotropic-case-2}
\end{align}
\end{itemize}
In this case, $\omega_{\pm} > 0$ which guarantees propagation of physical modes for all values of $k$. The phase and group velocities are equal,
\begin{align}
v_{\mathrm{ph} (\pm)} &\equiv \frac{\omega_{\pm}}{k} = \frac{1}{\sqrt{\mu\epsilon+\mu \alpha''^{2}} \mp \mu\alpha''}, \label{eq:group-velocity-isotropic-case-3} \\
 v_{\mathrm{g}(\pm)} &\equiv \left| \frac{\partial \omega_{\pm}}{\partial {\bf{k}}} \right| = \frac{1}{\sqrt{\mu\epsilon+\mu \alpha''^{2}} \mp \mu \alpha''} . \label{eq:group-velocity-isotropic-case-4}
\end{align}
\begin{itemize}
\item For $\alpha$, $\beta \in \mathbb{R}$, $(\alpha+\beta)=Z=0$, one has
\begin{align}
\omega_{\pm} &= \frac{k}{\sqrt{\mu\epsilon}}, \label{eq:group-velocity-isotropic-case-5}
\end{align}
\end{itemize}
which yields
\begin{align}
v_{\mathrm{ph}(\pm)}= v_{\mathrm{g}(\pm)}= \frac{1}{\sqrt{\mu\epsilon}} . \label{eq:group-velocity-isotropic-case-6}
\end{align}
Since $v_{\mathrm{g}(\pm)}<1$ in both cases $i)$ and $ii)$, the classical causality is ensured {for all $k$ and any value of $\alpha''$}.
}

{To examine the energy flux propagation direction, we implement Faraday's law and the constitutive relation (\ref{constitutive-iso2D}) {in the Poynting vector (\ref{eq:poynting-theorem-2}),} yielding
\begin{align}
{\bf{S}} &= \frac{1}{2\mu} \left[ {\bf{n}} |{\bf{E}}|^{2} - ({\bf{n}}\cdot {\bf{E}}) {\bf{E}}^{*} \right] + \frac{\beta^{*}}{2} ({\bf{E}}\times {\bf{E}}^{*} ) . \label{eq:poynting-vector-isotropic-1}
\end{align}
{In the absence of sources, the Gauss law (${\bf{k}}\cdot {\bf{D}}=0$), taking into account Eq. (\ref{DDiso}), provides ${\bf{k}}\cdot {\bf{E}} =0$}. Thus Eq.~\eqref{eq:poynting-vector-isotropic-1} is rewritten as
\begin{align}
{\bf{S}} &= \frac{1}{2\mu} {\bf{n}} |{\bf{E}}|^{2} + \frac{\beta^{*}}{2} ({\bf{E}}\times {\bf{E}}^{*}). \label{eq:poynting-vector-isotropic-2}
\end{align}
The real part of Eq.~(\ref{eq:poynting-vector-isotropic-2}) {provides} the time-averaged Poynting vector, that is,
\begin{align}
\left\langle {\bf{S}} \right\rangle &= \frac{1}{2\mu} {\bf{n}} |{\bf{E}}|^{2} + \mathrm{Re} \left[ \frac{\beta^{*}}{2} ({\bf{E}}\times {\bf{E}}^{*}) \right], \label{eq:poynting-vector-isotropic-3}
\end{align}
where we have used $\mathrm{Re}[{\bf{n}}]={\bf{n}}= n \hat{\bf{n}}$, since the refractive indices are real. Using the property $\mathrm{Re}[z] = (z+z^{*})/2$, with a complex $z$, ${\bf{E}}= {\bf{E}}' + \mathrm{i} {\bf{E}}''$, and $\beta= - \alpha^{*}$, the simplified Poynting vector takes the form
\begin{align}
\left\langle {\bf{S}} \right\rangle &= \frac{1}{2\mu} {\bf{n}} |{\bf{E}}|^{2} - \alpha'' ({\bf{E}}' \times {\bf{E}}'') . \label{eq:poynting-vector-isotropic-4}
\end{align}}
{The Gauss law, {${\bf{n}}\cdot {\bf{E}} =0$, implies ${\bf{n}}\cdot{\bf{E}}'=0$ and  ${\bf{n}}~\cdot~{\bf{E}}''=0$,} so both vectors ${\bf{E}}'$ and ${\bf{E}}''$ are in the plane orthogonal to $\bf{n}$. This way, the product ${\bf{E}}' \times {\bf{E}}''$ is parallel or antiparallel to $\bf{n}$.} {Therefore, in this bi-isotropic medium, the energy flux propagates along the same direction of the electromagnetic wave, independently of the value of the magnetoelectric parameter, $\alpha''$, responsible for the birefringence. }

{\section{Bi-anisotropic case with symmetric parameters\label{bianisotropic_Symm}}}

Now we explore the scenario where $\alpha _{ij}$ and $\beta _{ij}$ are
nondiagonal symmetric tensors, while the electric
permittivity, $\epsilon $, and the magnetic permeability, $\mu $, are simple
numbers. They can be easily parametrized by using a single 3-vector $%
\mathbf{d}$, that is,
\begin{equation}
\alpha _{ij}=\tilde{\alpha}d_{i}d_{j},\quad \beta _{ij}=\tilde{\beta}%
d_{i}d_{j},  \label{symmetric1}
\end{equation}%
in such a way that the constitutive relations take the form
\begin{subequations}
\label{constitutive-relations-symmetric-1}
\begin{align}
\mathbf{D}& =\epsilon \mathbf{E}+\tilde{\alpha}\mathbf{d}(\mathbf{d}\cdot
\mathbf{B}),  \label{symmetric2-1} \\
\mathbf{H}& =\frac{1}{\mu }\mathbf{B}+\tilde{\beta}\mathbf{d}(\mathbf{d}%
\cdot \mathbf{E}).  \label{symmetric2}
\end{align}%
\end{subequations}
The parameters $\alpha _{ij}$ and $\beta _{ij}$ in Eqs.~\eqref{symmetric1}
represent symmetric matrices with trace given by ${\tilde{\alpha}\mathbf{d}%
^{2}}$ and ${\tilde{\beta}\mathbf{d}^{2}}$, respectively. These $3\times 3$
matrices contain off-diagonal elements that could yield anisotropy. This is
a bold difference in relation to the bi-isotropic configuration
\eqref{isot5}, examined in Sec. \ref{bi-isotropic-case}. For a matter of
generality, we suppose $\alpha _{ij},\beta _{ij}\in \mathbb{C}$, which is
compatible with $\mathbf{d}\in \mathbb{R}^{3}$ and $\tilde{\alpha},\tilde{%
\beta}\in \mathbb{C}$. This way, in accordance with condition (\ref{relation-parameter3}), the parameters \eqref{symmetric1} should obey
\begin{equation}
\tilde{\beta}=-\tilde{\alpha}^{*},
\label{condition-for-symmetric-1}
\end{equation}
which implies
\begin{equation} \tilde{\alpha}'=-\tilde{\beta}', \quad \tilde{\alpha}''=\tilde{\beta}'',
\label{alphatildabetatilda}
\end{equation}
for $\tilde{\alpha}'=\mathrm{Re}[\tilde{\alpha}]$,
$\tilde{\alpha}''=\mathrm{Im}[\tilde{\alpha}]$,
$\tilde{\beta}'=\mathrm{Re}[\tilde{\beta}]$ and
$\tilde{\beta}''=\mathrm{Im}[\tilde{\beta}]$.

The constitutive relation (\ref{symmetric2-1}) allows us to write the
displacement vector in the form
\begin{equation}
\mathbf{D}=\epsilon \mathbf{E}+\frac{\tilde{\alpha}}{\omega }\mathbf{d}\left[
\mathbf{d}\cdot (\mathbf{k}\times \mathbf{E})\right] .
\end{equation}%
The Gauss law, $\mathbf{k}\cdot \mathbf{D}=0$, requires electric field configurations satisfying
\begin{equation}
\left[ \epsilon \mathbf{n}+\tilde{\alpha}(\mathbf{n}\cdot \mathbf{d})(%
\mathbf{d}\times \mathbf{n})\right] \cdot \mathbf{E}=0,  \label{gauss2cc}
\end{equation}%
for $\mathbf{k}=\omega \mathbf{n}$.

{
In the case $\mathbf{d}$ and $\mathbf{n}$ are parallel vectors, Eq.\eqref{gauss2cc} implies transversal electric field modes, $\mathbf{n} \cdot \mathbf{E}=0$. For nonparallel vectors $\mathbf{d}$ and $\mathbf{n}$, the electric field may be written as in Eq. \eqref{symmetric_arb} of Appendix A, which does not supply, in general, transversal modes, that is, $\mathbf{n}\cdot\mathbf{E}\neq 0$; see Eq. (\ref{symmetric_arbx}). Furthermore, for orthogonal vectors, ${\mathbf{n}\cdot\mathbf{d}=0}$, Eq. (\ref{gauss2cc}) yields again transversal {modes, ${\mathbf{n}}\cdot {\mathbf{E}}=0$,} with the electric field  expressed as in Eq. (\ref{symmetric_orthogonalx}).}

Replacing relations (\ref{symmetric1}) in the {permittivity tensor (\ref{ex7}),} ones writes
	\begin{align}  \label{symmetric3}
	\bar{\epsilon}_{ij}=\epsilon\delta_{ij}-\frac{1}{\omega}\left( \tilde{\beta%
	}   {\varepsilon_{imn}}        k_{m}d_{n}d_{j}+ \tilde{\alpha}    {\varepsilon_{amj}}   d_{i}d_{a}k_{m}%
	\right),
	\end{align}
	in such a way the tensor $M_{ij}$, Eq.~\eqref{ex12}, provides
	\begin{align}
	{M} = \mathcal{N} - \mu (\mathcal{D} + \mathcal{E}) , \label{symmetric4}
	\end{align}
	with $\mathcal{N}$ of Eq.\eqref{Nij1}, containing the usual constitutive elements, while the magnetoelectric  contributions are displayed in the following:
\begin{subequations}
\label{compact-matrices-symmetric-case-1}
	\begin{align}
	\mathcal{D} &= -(\tilde{\alpha}-\tilde{\beta})  \mathrm{diag} \left(D_{1}, D_{2}, D_{3} \right),  \\
		\mathcal{E} &= \begin{pmatrix}
	0 & \epsilon_{12} & \epsilon_{13} \\
	\epsilon_{21} & 0 & \epsilon_{23} \\
	\epsilon_{31} & \epsilon_{32} & 0
	\end{pmatrix} ,
	\end{align}
	where
	\begin{align}
	D_{1}&= d_{1} (d_{2} n_{3} -d_{3} n_{2}), \\
	D_{2}&=  d_{2} (d_{3} n_{1} - d_{1} n_{3}), \\
	D_{3} &=  d_{3} (d_{1} n_{2} - d_{2} n_{1} ) ,
	\end{align}
	\end{subequations}
and
\begin{subequations}
		\label{matrix-symmetric-case-extra-0}
		\begin{align}
		\epsilon_{12} &= -\tilde{\beta}d_{2} (d_{3}n_{2}-d_{2}n_{3}) + \tilde{\alpha}%
		d_{1}(d_{1}n_{3}-d_{3}n_{1}) ,  \label{symmetric-case-extra-1} \\
		\epsilon_{13} &= -\tilde{\beta}d_{3} (d_{3}n_{2} - d_{2}n_{3}) + \tilde{%
			\alpha} d_{1} (d_{2}n_{1}-d_{1}n_{2}) ,  \label{symmetric-case-extra-2} \\
		\epsilon_{21} &= -\tilde{\beta}d_{1} (d_{1}n_{3} -d_{3}n_{1}) + \tilde{\alpha%
		}d_{2} (d_{3}n_{2} - d_{2}n_{3}),  \label{symmetric-case-extra-3} \\
		\epsilon_{23} &= - \tilde{\beta} d_{3} (d_{1}n_{3} - d_{3}n_{1}) + \tilde{%
			\alpha}d_{2} (d_{2}n_{1} - d_{1}n_{2}) ,  \label{symmetric-case-extra-4} \\
		\epsilon_{31} &= - \tilde{\beta}d_{1} (d_{2} n_{1} - d_{1} n_{2} ) + \tilde{%
			\alpha} d_{3} (d_{3} n_{2} -d_{2}n_{3}),  \label{symmetric-case-extra-5} \\
		\epsilon_{32} &= -\tilde{\beta} d_{2} (d_{2} n_{1} - d_{1}n_{2}) + \tilde{%
			\alpha} d_{3} (d_{1}n_{3} - d_{3} n_{1} ) .  \label{symmetric-case-extra-6}
		\end{align}
			Evaluating $\mathrm{det}[M_{ij}]=0$, we obtain the dispersion relation,
	\end{subequations}
	\begin{align}  \label{symmetric5}
	{\epsilon} \left(n^{2}-\mu\epsilon \right)^{2} +
	\tilde{\alpha} \tilde{\beta} \mu \left[ \mu {\epsilon} d^{2} - (%
	\mathbf{n}\cdot \mathbf{d} )^{2} \right]
	 {\left( \mathbf{d}\times \mathbf{n}\right) ^{2}} =0 ,
	\end{align}
{ where $\left( \mathbf{d}\times \mathbf{n}\right) ^{2}\equiv d^2\mathbf{n}^2 -\left( \mathbf{d}\cdot\mathbf{n}\right) ^{2}$.}

Relation (\ref{condition-for-symmetric-1}) provides $\tilde{\alpha}%
\tilde{\beta}=-|\tilde{\alpha}|^{2}$.
Furthermore, implementing $\mathbf{n}\cdot \mathbf{d}=n
d\cos \varphi$, Eq.\eqref{symmetric5} yields
\begin{equation}
n_{\pm }^{2}=\frac{1}{s}\left[ N\pm \mu |\tilde{\alpha}|d^{2}\sin
^{2}\varphi \sqrt{\mu \epsilon +\frac{\mu ^{2}|\tilde{\alpha}|^{2}d^{4}}{4}}%
\right] ,  \label{refractive-indices-symmetric-1}
\end{equation}%
or
\begin{equation}
n_{\pm }=\sqrt{\frac{N+\mu \epsilon \sqrt{s}}{2s}}\pm \sqrt{%
\frac{N-\mu \epsilon \sqrt{s}}{2s}},  \label{refractive-indices-symmetric-1b}
\end{equation}%
where
\begin{subequations}
\label{refractive-indices-symmetric-2}
\begin{align}
N& =\mu \epsilon +\frac{\mu ^{2}|\tilde{\alpha}|^{2}d^{4}}{2}\sin
^{2}\varphi .  \label{refractive-indices-symmetric-4} \\
s& =1+\frac{\mu }{\epsilon }|\tilde{\alpha}|^{2}d^{4}\sin ^{2}\varphi \cos
^{2}\varphi .  \label{refractive-indices-symmetric-3}
\end{align}
\end{subequations}

We notice in Eq.\eqref{refractive-indices-symmetric-1b} two distinct refractive
indices, both real and positive, $n_{\pm}>0$,
in such a way that birefringent electromagnetic propagation is expected in this medium. {We also highlight that Eq.~\eqref{refractive-indices-symmetric-1} holds for any propagation direction in relation to the vector $\mathbf{d}$, generally expressed in terms of the angle $\varphi$, {the relative angle between the vector ${\bf{d}}$ and the propagation direction.}}

\begin{figure}[H]
\begin{centering}
\includegraphics[scale=.5]{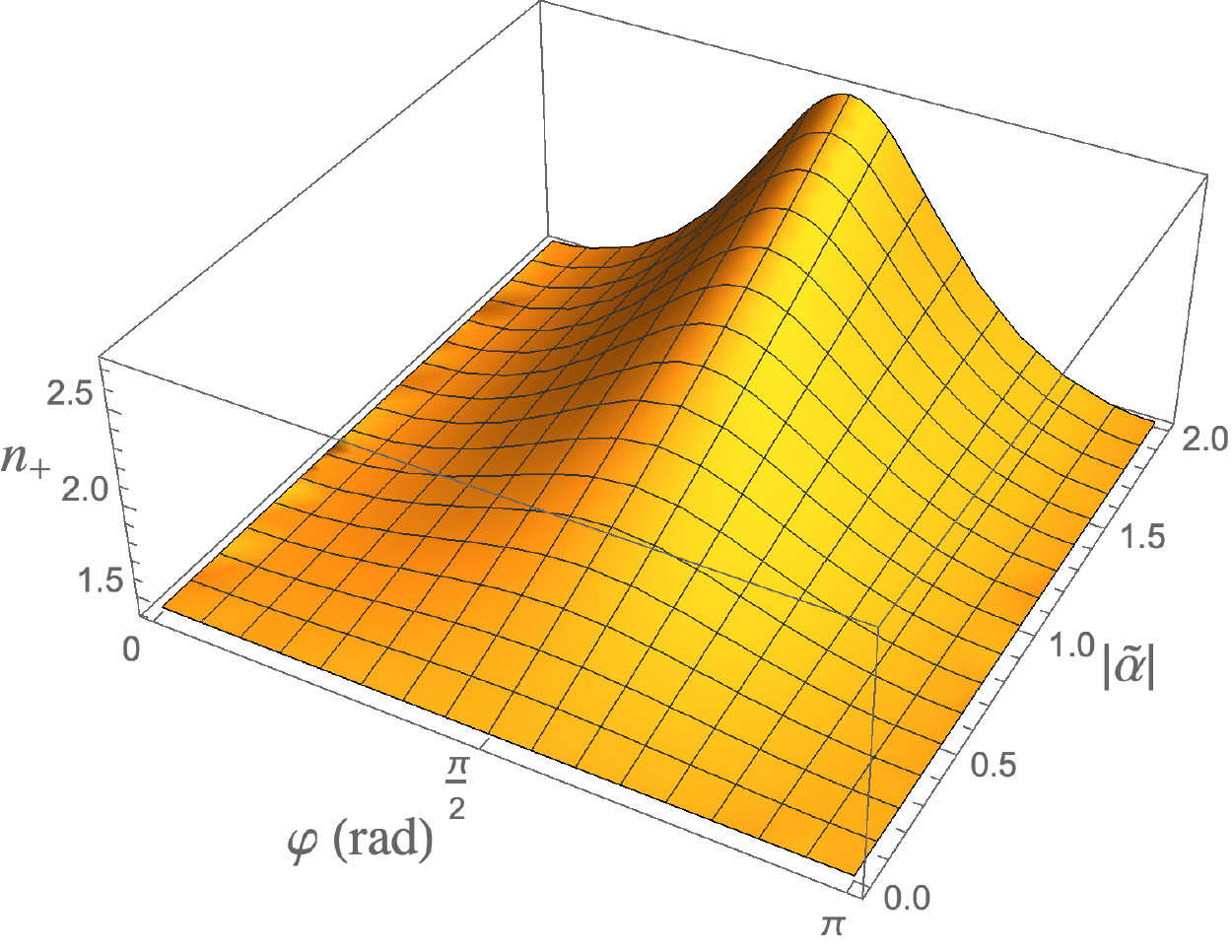}
\par\end{centering}
\caption{\label{plot3D-indice-de-refracao-mais-caso-simetrico} Refractive index $n_{+}$ of  Eq.~\eqref{refractive-indices-symmetric-1b} with $\mu=1$, $\epsilon=2$, and $d=1$. The parameters $\epsilon$, $\mu$, and $|\tilde{\alpha}|d^{2}$ are dimensionless.}
\end{figure}

\begin{figure}[H]
\begin{centering}
\includegraphics[scale=.52]{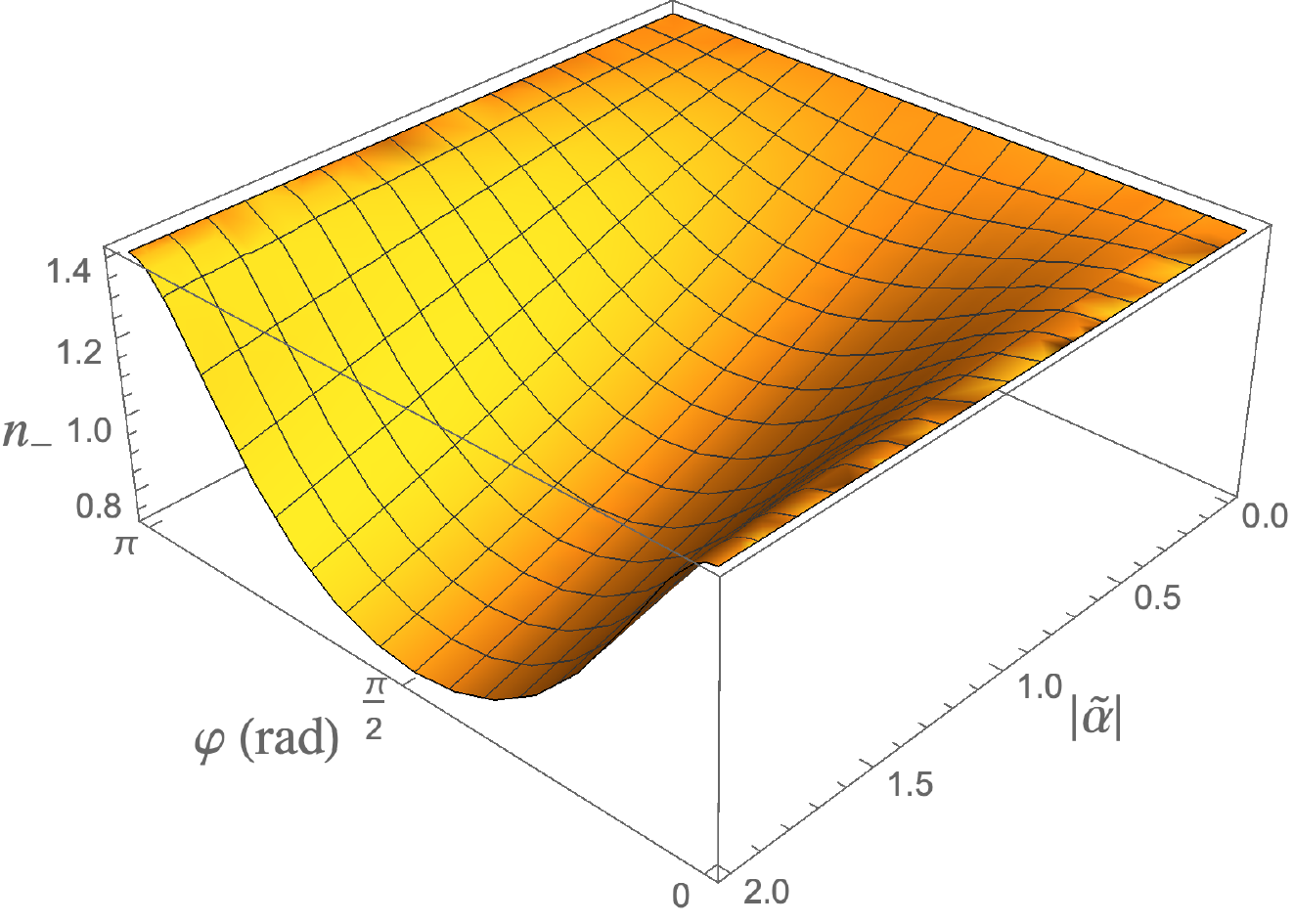}
\par\end{centering}
\caption{\label{plot3D-indice-de-refracao-menos-caso-simetrico-tipo-2} Refractive index $n_{-}$ of  Eq.~\eqref{refractive-indices-symmetric-1b} with $\mu=1$, $\epsilon=2$, and $d=1$. {The parameters $\epsilon$, $\mu$, and $|\tilde{\alpha}|d^{2}$ are dimensionless.}}
\end{figure}

{
The behavior of $n_{\pm}$ [Eq.~(\ref{refractive-indices-symmetric-1b})] in terms of $\varphi \in [0, \pi]$ and {the dimensionless parameter $|\tilde{\alpha}| \in [0,1]$} is illustrated in Figs.~\ref{plot3D-indice-de-refracao-mais-caso-simetrico} and \ref{plot3D-indice-de-refracao-menos-caso-simetrico-tipo-2}.} {
The anisotropy effect manifests itself by means of the angular dependence of $n_{\pm}$ on $\varphi$.  Notice the nonlinearity of $n_{\pm}$, which behaves as a sinusoidal function, increasing with $|\tilde{\alpha}|$. }

In the following, we address the propagating modes and examine birefringence effects {for some specific propagation directions.}

{\subsection{\label{section-propagation-modes-symmetric-case}Propagation modes}}

To examine the propagating modes, we first pay attention to the
angle between $\mathbf{n}$ and $\mathbf{d}$. For the configurations where $%
\mathbf{n}$ and $\mathbf{d}$ are in the same direction,
Eq.\eqref{refractive-indices-symmetric-1} yields $n_{\pm}=\sqrt{\mu\epsilon}$,
which is the usual scenario. Modified scenarios arise when $\mathbf{n}$ and $%
\mathbf{d}$ are not aligned. To obtain the modes, we choose a simplified
coordinate system where the vector $\mathbf{n}$ is along the $z$-axis, that
is, $\mathbf{n}=(0,0, n)$. For such a choice, we investigate
a configuration where the background vector is longitudinal and orthogonal
to $\mathbf{n}$.

\bigskip

\subsubsection{$\mathbf{d}$-Longitudinal configuration:}

We begin examining the $\mathbf{d}$-longitudinal configuration,
\begin{equation}
\mathbf{d}=(0,0,d),  \label{d-longitudinal}
\end{equation}%
for which Eq.\eqref{refractive-indices-symmetric-1} yields
\begin{equation}
n^{2}=\mu \epsilon .
\end{equation}%
For this case, matrix (\ref{symmetric4}) is rewritten as
\begin{equation}
{M}=%
\begin{pmatrix}
n^{2}-\mu \epsilon & 0 & 0 \\
0 & n^{2}-\mu \epsilon & 0 \\
0 & 0 & -\mu \epsilon%
\end{pmatrix}%
,  \label{propagating-symmetric-4}
\end{equation}%
in such a way that $M_{ij}E^{j}=0$ provides generic orthogonal modes
\begin{equation}
\mathbf{E}=%
\begin{pmatrix}
E_{x} \\
E_{y} \\
0%
\end{pmatrix}%
,  \label{propagating-symmetric-6}
\end{equation}%
representing a transversal wave with undefined polarization (linear, circular, or elliptical). It is interesting to note that only one positive refractive index was determined and it does not depend on the propagation direction, which is a signal of isotropy. This means that the
${\bf{d}}$-direction defines the optical axis of the medium.

\subsubsection{$\mathbf{d}$-transversal configuration \label{anti_bb}}

We follow writing the $\mathbf{d}$-transversal configuration,
\begin{equation}
\mathbf{d}=(d_{1},d_{2},0),  \label{d-T}
\end{equation}%
for which $s=1$ and $N=\mu {\epsilon }+|\tilde{\alpha}|^{2}\mu ^{2}d^{4}/2$,
so that Eq.~\eqref{refractive-indices-symmetric-1} yields
\begin{equation}
n_{\pm }^{2}=\mu \epsilon +\frac{\mu ^{2}|\tilde{\alpha}|^{2}d^{4}}{2}\pm
\mu |\tilde{\alpha}|d^{2}\sqrt{\mu \epsilon +\frac{\mu ^{2}|\tilde{\alpha}%
|^{2}d^{4}}{4}},  \label{symmetric6T}
\end{equation}%
and
\begin{equation}
n_{\pm }=\sqrt{\mu \epsilon +\frac{\mu ^{2}|\tilde{\alpha}|^{2}d^{4}}{4}}\pm
\frac{\mu |\tilde{\alpha}|d^{2}}{2}.  \label{symmetric6T2}
\end{equation}
By using Eq. (\ref{symmetric6T2}) we rewrite Eq. (\ref%
{symmetric6T}) as
\begin{equation}
n_{\pm }^{2}=\mu \epsilon \pm \mu |\tilde{\alpha}| d^{2} n_\pm.
\label{symmetric6T0}
\end{equation}%

Matrix (\ref{symmetric4}) now is
\begin{equation}
{M}=%
\begin{pmatrix}
n_{\pm }^{2}-\mu \epsilon +\Omega n_{\pm } & -\mu n_{\pm }(\tilde{\beta}%
d_{2}^{2}+\tilde{\alpha}d_{1}^{2}) & 0 \\
\mu n_\pm(\tilde{\beta}d_{1}^{2}+ \tilde{\alpha}d_{2}^{2})
 & n_{\pm }^{2}-\mu \epsilon -\Omega n_{\pm } & 0 \\
0 & 0 & -\mu \epsilon%
\end{pmatrix} ,
\end{equation}
or better,
\begin{equation}
{M}=
\begin{pmatrix}
(\pm \mu |\tilde{\alpha}| d^{2}+\Omega )n_{\pm } & -\mu n_{\pm
}(\tilde{\beta}d_{2}^{2}+\tilde{\alpha}d_{1}^{2}) & 0 \\
\mu n_{\pm }(\tilde{\beta}d_{1}^{2}+\tilde{\alpha}d_{2}^{2}) & (
\pm \mu |\tilde{\alpha}| d^{2}-\Omega )n_{\pm } & 0 \\
0 & 0 & -\mu \epsilon%
\end{pmatrix}%
,
\end{equation}
where we have used Eqs.\eqref{alphatildabetatilda} and \eqref{symmetric6T0},
writing
\begin{align}
\Omega & =\mu (\tilde{\alpha}-\tilde{\beta})d_{1}d_{2}=2\mu \tilde{\alpha}%
^{\prime }d_{1}d_{2}.
\end{align}%
The condition $M_{ij}E^{j}=0$ yields
\begin{equation}
\mathbf{E}_{\pm }=E_{0}%
\begin{pmatrix}
1 \\
\displaystyle\frac{\mu \left( \tilde{\beta}d_{1}^{2}+\tilde{\alpha}%
d_{2}^{2}\right) }{\Omega \mp \mu |\tilde{\alpha}|d^{2}} \\
0%
\end{pmatrix},
\end{equation}%
with an appropriately chosen amplitude $E_{0}$.
For
$d_{1}=0$, we achieve
\begin{equation}
\mathbf{E}_{\pm }=E_{0}%
\begin{pmatrix}
1 \\
\displaystyle\mp \frac{\tilde{\alpha}}{|\tilde{\alpha}|} \\
0%
\end{pmatrix}%
=\frac{1}{\sqrt{2}}%
\begin{pmatrix}
1 \\
\displaystyle\mp \frac{\tilde{\alpha}^{\prime }+\mathrm{i}\tilde{\alpha}%
^{\prime \prime }}{|\tilde{\alpha}|} \\
0%
\end{pmatrix}%
,  \label{propagating-symmetric-6B-1}
\end{equation}%
which represents linear polarizations for $\tilde{\alpha}^{\prime \prime
}=0$ or circular polarizations for $\tilde{\alpha}^{\prime }=0$.

As Eq.\eqref{symmetric6T2} exhibits two real refractive indices,
a scenario with birefringence is set. The modes
\eqref{propagating-symmetric-6B-1}, however, do not represent RCP or LCP
vectors, so the birefringent propagation cannot be suitably described
in terms of the rotatory power \eqref{eq:rotatory-power1A}.
Rather, it can be characterized in terms of the phase
shift arising from the distinct phase velocities of the propagating modes,
given by
\begin{equation}
\Delta =\frac{2\pi }{\lambda _{0}}l(n_{+}-n_{-}),  \label{phase-shift-0}
\end{equation}%
where $\lambda _{0}$ is the vacuum wavelength of incident light, $l$ is the thickness of the medium or the distance
traveled by the wave, and $n_{+}$ and $n_{-}$	are the refractive indices of the medium. Note that this is the same expression that controls the phase shift caused
by ``retarders'' (for details, see Chap. 8 of Ref. \cite{Hecht}). Using %
Eq.\eqref{symmetric6T2}, one finds the corresponding phase shift per unit
length as
\begin{equation}
\frac{\Delta }{l}=\frac{2\pi }{\lambda _{0}}\mu |\tilde{\alpha}|d^{2}.
\label{phase-shift-1}
\end{equation}
As the phase shift depends on the modulus of $\tilde{\alpha}$, the birefringence now takes place for both real and imaginary magnetoelectric parameters. This is a difference in relation to the bi-isotropic case \eqref{constitutive-relations-biisotropic-1}, in which the birefringence occurs only for imaginary parameters, as shown in Eq.\eqref{isotropic-case-1-3}.

\subsubsection{$\mathbf{d}$-general configuration\label{anti_cc}}

Now, let us analyze the mixed case where the vector $%
\mathbf{d}$ has orthogonal and longitudinal components relative to the
propagation direction, $\mathbf{n}$. In this sense, one can set
\begin{equation}
\mathbf{d}=(0,d_{2},d_{3}).  \label{symmetric-mixed-case-1}
\end{equation}%
The refractive indices
(\ref{refractive-indices-symmetric-1}) are rewritten as
\begin{equation}
n_{\pm }^{2}=\frac{1}{s}\left( \mu \epsilon +\Lambda _{\pm
}\sin ^{2}\varphi \right) ,  \label{symmetric-mixed-case-2}
\end{equation}
where $s$ is given by Eq.~\eqref{refractive-indices-symmetric-3}
and $\Lambda_{\pm}$ is defined as
\begin{equation}
\Lambda_{\pm}= \frac{\mu^{2}|\tilde{\alpha}|^{2} d^{4}}{2} \pm \mu |\tilde{%
\alpha}| d^{2} \sqrt{\mu\epsilon + \frac{\mu^{2} |\tilde{\alpha}|^{2}d^{4}}{4%
}}.
\end{equation}%

For the coordinate system where $\mathbf{n}=(0,0,n_{3})$ and $\mathbf{d}$ is
given by Eq.\eqref{symmetric-mixed-case-1}, matrix \eqref{symmetric4} takes the form
\begin{equation}
{M}=%
\begin{pmatrix}
n_{3}^{2}-\mu \epsilon & -\mu \tilde{\beta}d_{2}^{2}n_{3} &
-\mu \tilde{\beta}d_{2}d_{3}n_{3} \\
+\mu \tilde{\alpha}d_{2}^{2}n_{3} & n_{3}^{2}-\mu \epsilon & 0 \\
+\mu \tilde{\alpha}d_{2}d_{3}n_{3} & 0 & -\mu \epsilon%
\end{pmatrix}%
,
\end{equation}%
which becomes
\begin{equation}
{M}=%
\begin{pmatrix}
1/\Upsilon _{\pm } & -\mu \tilde{\beta}d_{2}^{2}n_{3} &
-\mu \tilde{\beta}d_{2}d_{3}n_{3} \\
&  &  \\
+\mu \tilde{\alpha}d_{2}^{2}n_{3} & 1/\Upsilon _{\pm } & 0 \\
&  &  \\
+\mu \tilde{\alpha}d_{2}d_{3}n_{3} & 0 & -\mu \epsilon%
\end{pmatrix}%
,
\end{equation}%
when Eq.\eqref{symmetric-mixed-case-2} is taken into account
and we have defined
\begin{equation}
\Upsilon _{\pm }=\frac{s}{\mu \epsilon \left( 1-s\right) +\Lambda _{\pm }\sin ^{2}\varphi
}. \label{Upssilon}
\end{equation}%
The condition $M_{ij}E^{j}=0$ yields
\begin{equation}
\mathbf{E}_{\pm }=E_{0}%
\begin{pmatrix}
1 \\
-\mu \left( \tilde{\alpha}^{\prime } +\mathrm{i}\tilde{\alpha}%
^{\prime \prime} \right) d_{2}^{2}n_{\pm }\Upsilon _{\pm } \\
+\left( \tilde{\alpha}^{\prime }+\mathrm{i}\tilde{\alpha}%
^{\prime \prime }\right) d_{2}d_{3}n_{\pm }/\epsilon%
\end{pmatrix}%
,  \label{symmetric-mixed-case-6}
\end{equation}
where $n_{\pm}$ represents $n_{3\pm}$.
Differently from the previous $\mathbf{d}$-longitudinal or
${\bf{d}}$-orthogonal cases, modes (\ref{symmetric-mixed-case-6}) are
endowed with a longitudinal component, a feature of general solutions for
{$\varphi\neq 0,\pi$} or $\varphi\neq \pi/2$.

\begin{figure}[h!]
\begin{centering}
\includegraphics[scale=.6]{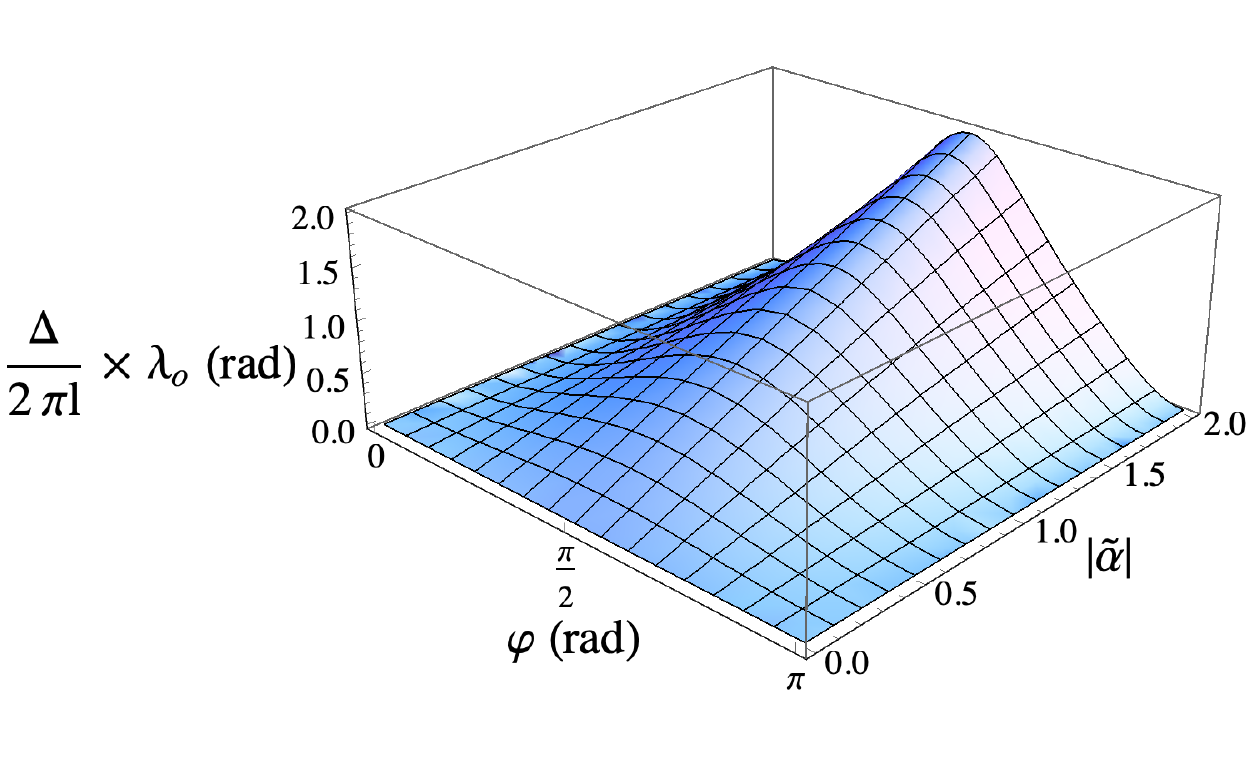}
\par\end{centering}
\caption{\label{phase-shift-extended-symmetric-general-case} Phase shift factor per unit length of Eq.~\eqref{symmetric-mixed-case-8}. Here, we have used $\mu=1$, $\epsilon=2$, and $d=1$.}
\end{figure}

The polarization of modes (\ref{symmetric-mixed-case-6}) can be read off their transversal sectors. Since the transversal piece of Eq.(\ref{symmetric-mixed-case-6}) is neither RCP nor LCP, being linearly (for $\tilde{\alpha}^{\prime\prime }=0$) or elliptically polarized (for $\tilde{\alpha}^{\prime }=0$), the birefringence effect is expressed by means of the phase shift per unit length defined in Eq.\eqref{phase-shift-0}, here carried out as
	\begin{equation}
	\label{symmetric-mixed-case-8}
	\frac{\Delta}{l} = \frac{4\pi}{\lambda_{0}}  \sqrt{\frac{N-\mu \epsilon \sqrt{s}}{2s}},
	\end{equation}
where we used  Eq. (\ref{refractive-indices-symmetric-1b}). {In Fig. \ref{phase-shift-extended-symmetric-general-case} we have plotted the phase shift (\ref{symmetric-mixed-case-8}) per unit length in terms of $\varphi$ and $|\tilde{\alpha}|$. We notice that the birefringence effect {is maximal} for $\varphi=\pi/2$, which corresponds to configurations where $n_{+}$ and $n_{-}$ have maxima and minima values, respectively. }

As a final comment, note that for $d_{2}=0$ or $\mathbf{d}=(0,0,d_{3})$  the field (\ref{symmetric-mixed-case-6}) simplifies as 	
	\begin{equation}
	\mathbf{E}_{\pm }=%
	\begin{pmatrix}
	1 \\
	0 \\
	0 \\
	\end{pmatrix},
 \label{symmetric-mixed-case-6B}
	\end{equation}
which corresponds to the \textbf{d}-longitudinal case, whose mode is given by Eq.\eqref{propagating-symmetric-6}, being compatible with the result (\ref{symmetric-mixed-case-6B}). On the other hand, for the transversal configuration, $d_{3}=0$ or $\mathbf{d}=(0,d_{2},0)$, the field (\ref{symmetric-mixed-case-6}) yields
	\begin{equation}
	\mathbf{E}_{\pm }=E_{0}%
	\begin{pmatrix}
	1 \\
	-\mu \left( \tilde{\alpha}^{\prime } +\mathrm{i}\tilde{\alpha}%
	^{\prime \prime} \right) d_{2}^{2}n_{\pm }\Upsilon _{\pm } \\
	0
	\end{pmatrix}%
	. \label{symmetric-mixed-case-6C}
	\end{equation}
	
For this transversal configuration  it holds that $\Upsilon _{\pm }=1/\Lambda_{\pm}$ and $\Lambda_{\pm}=\mu |\tilde{\alpha}| d^{2} n_{\pm }$.  With that, solution (\ref{symmetric-mixed-case-6C}) recovers the one of Eq.\eqref{propagating-symmetric-6B-1}.

{\subsection{Group velocity, phase velocity, and Poynting vector}}

\begin{figure}[H]
\begin{centering}
\includegraphics[scale=.6]{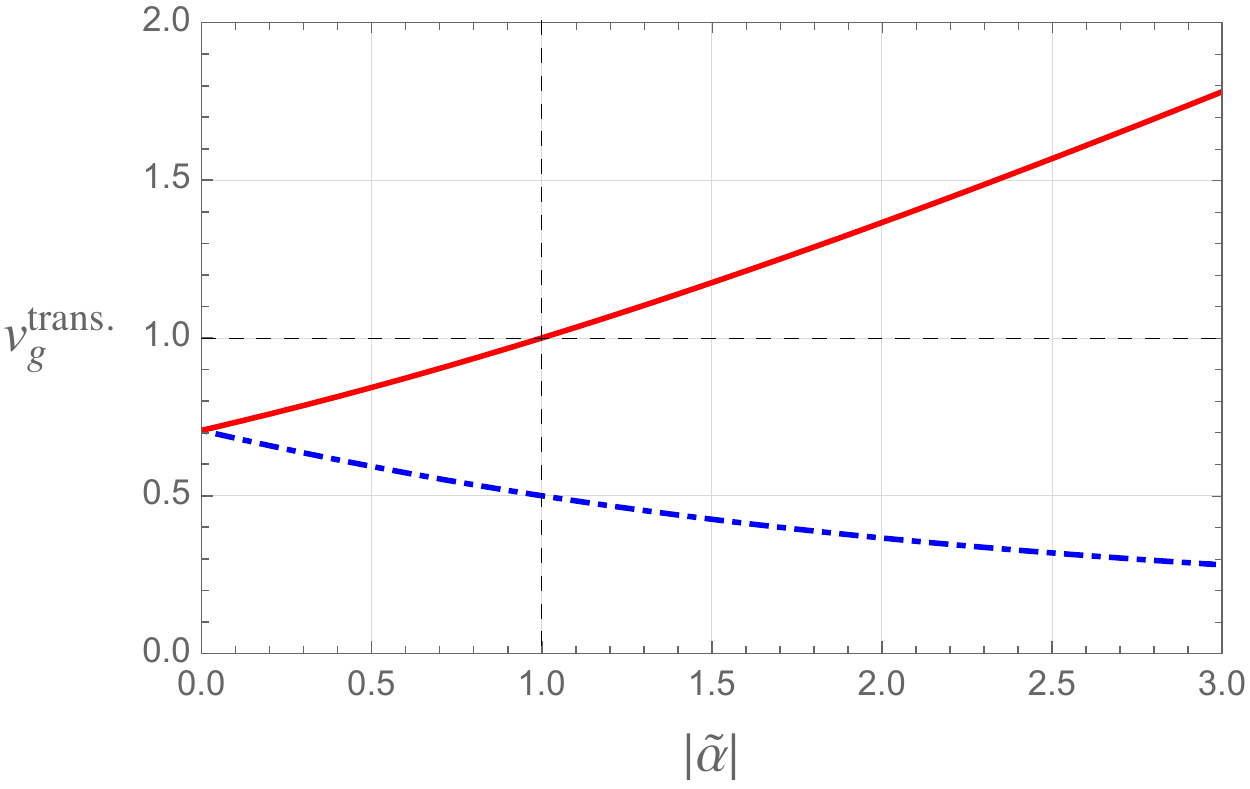}
\par\end{centering}
\caption{\label{plot-symmetric-case-trasnversal-group-veloctiy} Group velocity $v_{\mathrm{g}(\pm)}$ [Eq.~(\ref{eq:bi-anisotropic-symmetric-case-group-velocity-8})] of ${\bf{d}}$-transversal case. The blue dot-dashed curve indicates $v_{\mathrm{g}(+)}$, while the red solid line represents $v_{\mathrm{g}(-)}$. {The vertical dashed line indicates $|\tilde{\alpha}|=1$, corresponding to the value of the critical factor $\alpha_{c}$  [Eq.~\eqref{eq:bi-anisotropic-symmetric-case-group-velocity-9}] for the choices $\mu=1$, $\epsilon=2$ and $d=1$, the ones adopted in this plot.}}
\end{figure}

{The dispersion equation (\ref{symmetric5}) can be rewritten in the form
\begin{align}
\omega^{4} - 2\omega^{2} \left\{ \frac{ {\bf{k}}^{2}}{\mu\epsilon} + \frac{ |\tilde{\alpha}|^{2} d^{2}}{2\epsilon^{2}} {({\bf{d}}\times{\bf{k}})^{2}} \right\} + \nonumber \\
+ \frac{\mu}{\epsilon} \frac{ |\tilde{\alpha}|^{2}}{\mu^{2}\epsilon^{2}} ({\bf{k}}\cdot {\bf{d}})^{2} {({\bf{d}}\times{\bf{k}})^{2}}  +  \frac{ {\bf{k}}^{4}}{\mu^{2}\epsilon^{2}} &= 0 , \label{eq:bi-anisotropic-symmetric-case-group-velocity-1}
\end{align}
whose solutions for $\omega$,
\begin{align}
\omega_{\pm}^{2} &= \frac{ {\bf{k}}^{2}}{\mu\epsilon} + \frac{ |\tilde{\alpha}|}{\epsilon} {({\bf{d}}\times{\bf{k}})^{2}}  \left[ \frac{ |\tilde{\alpha}|d^{2}}{2\epsilon} {\mp} \sqrt{ \frac{1}{\mu\epsilon} + \frac{ |\tilde{\alpha}|^{2}d^{4}}{4\epsilon^{2}}} \right]  . \label{eq:bi-anisotropic-symmetric-case-group-velocity-2}
\end{align}
provide the phase and group velocities,
\begin{align}
v_{\mathrm{ph}(\pm)} \equiv \frac{\omega_{\pm}}{k} = \sqrt{ \frac{1}{\mu\epsilon} + \frac{ |\tilde{\alpha}|}{\epsilon} {({\bf{d}}\times{\hat{\bf{k}}})^{2}}  \left(\frac{ |\tilde{\alpha}| d^{2}}{2\epsilon} {\mp} f_{\alpha} \right) } , \label{eq:bi-anisotropic-symmetric-case-group-velocity-3}
\end{align}
\begin{align}
v_{g(\pm)}^{i} &= \frac{k^{i}}{\mu\epsilon \omega_{\pm}} + \frac{ |\tilde{\alpha}|}{\epsilon \omega_{\pm}} \left[ d^{2} k^{i} - ({\bf{d}}\cdot {\bf{k}}) d^{i} \right] \left( \frac{ |\tilde{\alpha}| d^{2}}{2\epsilon} {\mp} f_{\alpha} \right)  ,\label{eq:bi-anisotropic-symmetric-case-group-velocity-4}
\end{align}
with
\begin{align}
f_{\alpha} = \sqrt{\frac{1}{\mu\epsilon} + \frac{ |\tilde{\alpha}|^{2}d^{4}}{4\epsilon^{2}}}. \label{eq:bi-anisotropic-symmetric-case-group-velocity-5}
\end{align}
Both $v_{\mathrm{ph}(\pm)}$ and $v_{\mathrm{g}(\pm)}$ are valid for general configurations, i. e., for any relative orientation between the vectors ${\bf{d}}$ and ${\bf{n}}$, {for which the group velocity is no longer parallel to ${\bf{n}}$, due to its component along the ${\bf{d}}$ vector.}  Considering now the special cases discussed in Sec.~\ref{section-propagation-modes-symmetric-case}, we state the following:
\begin{itemize}
\item For the ${\bf{d}}$-longitudinal scenario where ${\bf{d}}\cdot \hat{\bf{k}}= d$, Eqs. (\ref{eq:bi-anisotropic-symmetric-case-group-velocity-3}) and (\ref{eq:bi-anisotropic-symmetric-case-group-velocity-4}) provides
\end{itemize}
{\begin{align}
v_{\mathrm{ph}(\pm)}^{\mathrm{long.}} = \frac{1}{\sqrt{\mu\epsilon}}, \quad {\bf{v}}_{\mathrm{g}(\pm)}^{\mathrm{long.}} = \frac{\hat{\bf{k}}}{\sqrt{\mu\epsilon}} . \label{eq:bi-anisotropic-symmetric-case-group-velocity-6}
\end{align}}
\begin{itemize}
\item For the ${\bf{d}}$-transversal case where ${\bf{d}}\cdot \hat{\bf{k}}=0$, one obtains
\end{itemize}
\begin{align}
v_{\mathrm{ph}(\pm)}^{\mathrm{trans.}} &= \sqrt{\frac{1}{\mu\epsilon} + \frac{ |\tilde{\alpha}|^{2} d^{4}}{4 \epsilon^{2}}} {\mp} \frac{ |\tilde{\alpha}| d^{2}}{2\epsilon} , \label{eq:bi-anisotropic-symmetric-case-group-velocity-7} \\
{\bf{v}}_{\mathrm{g}(\pm)}^{\mathrm{trans.}} &= \left[\sqrt{\frac{1}{\mu\epsilon} + \frac{ |\tilde{\alpha}|^{2} d^{4}}{4 \epsilon^{2}}} {\mp} \frac{ |\tilde{\alpha}| d^{2}}{2\epsilon} \right] \hat{\bf{k}}. \label{eq:bi-anisotropic-symmetric-case-group-velocity-8}
\end{align}
 {The magnitude of the group velocity  (\ref{eq:bi-anisotropic-symmetric-case-group-velocity-8}) is depicted in terms of the magnetoelectric parameter in} Fig.~\ref{plot-symmetric-case-trasnversal-group-veloctiy}.
}

{We note {that $v_{g(-)} >1$ occurs} for $|\tilde{\alpha}| >\alpha_{c}$, where the critical value $\alpha_{c}$ is given by
\begin{align}
\alpha_{c} =\frac{\mu\epsilon-1}{\mu d^{2}} . \label{eq:bi-anisotropic-symmetric-case-group-velocity-9}
\end{align}
Such a mode with $v_{\mathrm{g}(-)}>1$ is related to the refractive index $n_{-}$ of Eq.~\eqref{refractive-indices-symmetric-1b}, which can assume values $n<1$ (see Fig.~\ref{plot3D-indice-de-refracao-menos-caso-simetrico-tipo-2}).
}

{Let us now evaluate the Poynting vector. Using the constitutive relation (\ref{symmetric2}), we obtain
\begin{align}
{\bf{S}} &= \frac{1}{2\mu} {\bf{n}} |{\bf{E}}|^{2} -\frac{1}{2\mu} ({\bf{n}}\cdot {\bf{E}}) {\bf{E}}^{*} + \frac{\beta^{*}}{2} ({\bf{E}}\times {\bf{d}}) ({\bf{d}}\cdot {\bf{E}}^{*}) . \label{eq:poynting-vector-symmetric-case-1}
\end{align}
As the Gauss law yields $\epsilon {\bf{k}} \cdot {\bf{E}}= -\tilde{\alpha} ({\bf{k}}\cdot {\bf{d}}) ({\bf{d}}\cdot {\bf{B}})$, {Eq. \eqref{eq:poynting-vector-symmetric-case-1} takes the form}
\begin{align}
{\bf{S}} &= \frac{1}{2\mu} {\bf{n}} |{\bf{E}}|^{2} + \frac{\tilde{\alpha}}{2\mu\epsilon} ({\bf{n}}\cdot {\bf{d}}) \left[ {\bf{E}} \cdot ({\bf{d}}\times {\bf{n}} ) \right] {\bf{E}}^{*} + \nonumber \\
&\phantom{=}+ \frac{\tilde{\beta}^{*}}{2} ({\bf{E}}\times {\bf{d}} )   ({\bf{d}}\cdot {\bf{E}}^{*})  . \label{eq:poynting-vector-symmetric-case-2}
\end{align}
Considering $\tilde{\beta}=-\tilde{\alpha}^{*}$ and ${\bf{E}}={\bf{E}}' + \mathrm{i} {\bf{E}}''$, the time-averaged Poynting vector, obtained {from} the real part of Eq.~(\ref{eq:poynting-vector-symmetric-case-2}), is given by
\begin{align}
\left\langle {\bf{S}} \right\rangle &= \frac{1}{2\mu} {\bf{n}} |{\bf{E}}|^{2} + \frac{\tilde{\alpha}'}{2\mu\epsilon} ({\bf{n}}\cdot {\bf{d}}) \left\{ \frac{}{} {\bf{E}}' \left[ {\bf{E}}' \cdot ({\bf{d}}\times {\bf{n}}) \right] \right. + \nonumber \\
&\phantom{=}\left. + {\bf{E}}'' \left[ {\bf{E}}'' \cdot ({\bf{d}}\times {\bf{n}}) \right] \right\}  - \frac{\tilde{\alpha}''}{2\mu\epsilon} ({\bf{n}}\cdot {\bf{d}}) \left\{ \frac{}{} {\bf{E}}' \left[ {\bf{E}}'' \cdot ({\bf{d}}\times {\bf{n}}) \right] + \right. \nonumber \\
&\phantom{=} \left. -{\bf{E}}'' \left[ {\bf{E}}' \cdot ({\bf{d}}\times {\bf{n}}) \right] \right\} - \frac{\tilde{\alpha}'}{2} \left[ ({\bf{d}}\cdot {\bf{E}}')    ({\bf{E}}' \times {\bf{d}}) + \right. \nonumber \\
&\phantom{=} \left. + ({\bf{d}}\cdot {\bf{E}}'')   ({\bf{E}}''\times {\bf{d}}) \right] + \frac{\tilde{\alpha}''}{2} \left[ ({\bf{d}}\cdot {\bf{E}}')  ({\bf{E}}''\times {\bf{d}}) + \right. \nonumber \\
&\phantom{=} \left. - ({\bf{d}}\cdot {\bf{E}}'')       ({\bf{E}}' \times {\bf{d}}) \right] , \label{eq:poynting-vector-symmetric-case-3}
\end{align}
or, equivalently,
\begin{align}
\left\langle {\bf{S}} \right\rangle &= \frac{1}{2\mu} {\bf{n}} |{\bf{E}}|^{2} + \frac{ ({\bf{n}}\cdot {\bf{d}})}{2\mu\epsilon}  \left( \tilde{\alpha}' {\bf{f}}^{\prime} - \tilde{\alpha}'' {\bf{f}}^{\prime\prime}\right) + \nonumber \\
&\phantom{=}- \frac{1}{2} \left(\tilde{\alpha}' {\bf{g}}^{\prime} - \tilde{\alpha}'' {\bf{g}}^{\prime\prime}\right) , \label{eq:poynting-vector-symmetric-case-4}
\end{align}
where
\begin{align}
{\bf{f}}^{\prime,\prime\prime}&= {\bf{E}}^{\prime} \left[ {\bf{E}}^{\prime,\prime\prime} \cdot ({\bf{d}}\times {\bf{n}}) \right] \pm {\bf{E}}^{\prime\prime} \left[ {\bf{E}}^{\prime\prime,\prime} \cdot ({\bf{d}}\times {\bf{n}}) \right] , \label{eq:poynting-vector-symmetric-case-5} \\[0.2cm]
{\bf{g}}^{\prime, \prime\prime} &= ({\bf{d}}\cdot {\bf{E}}^{\prime} )   ({\bf{E}}^{\prime, \prime\prime} \times {\bf{d}}) \pm ({\bf{d}} \cdot {\bf{E}}^{\prime\prime})   ({\bf{E}}^{\prime\prime, \prime} \times {\bf{d}}) . \label{eq:poynting-vector-symmetric-case-6}
\end{align}}
{In the case $\mathbf{d}$ and $\mathbf{n}$ are parallel vectors, Eq. \eqref{gauss2cc} implies transversal electric field modes, $\mathbf{n} \cdot \mathbf{E}=0$, then $\mathbf{d} \cdot \mathbf{E}=0$. In this case, we have ${\bf{f}}^{\prime,\prime\prime}=0$ and ${\bf{g}}^{\prime, \prime\prime}=0$ yielding simply $\left\langle {\bf{S}} \right\rangle = {\bf{n}} |{\bf{E}}|^{2}/{2\mu}$, and the energy flux propagates along the same direction of  $\mathbf{n}$.}

{For the case that the vectors $\mathbf{d}$ and $\mathbf{n}$ are mutually orthogonal, $\mathbf{n}\cdot\mathbf{d}=0$, Eq. \eqref{gauss2cc} also provides $\mathbf{n}\cdot \mathbf{E}=0$, so that
\begin{align}
\left\langle {\bf{S}} \right\rangle &= \frac{1}{2\mu} {\bf{n}} |{\bf{E}}|^{2} - \frac{1}{2} \left(\tilde{\alpha}' {\bf{g}}^{\prime} - \tilde{\alpha}'' {\bf{g}}^{\prime\prime}\right) . \label{eq:poynting-vector-symmetric-case-4B}
\end{align}}
	
{This scenario is such that the vectors} $\mathbf{d}$, ${\bf{E}}^{\prime}$, and ${\bf{E}}^{\prime \prime}$ are in the same plane orthogonal to $\mathbf{n}$. {This way, the vectors ${\bf{E}}' \times {\bf{d}}$ and ${\bf{E}}'' \times {\bf{d}}$ are along the ${\bf{n}}$-direction or are both null (when ${\bf{E}}$ and ${\bf{d}}$ are parallel vectors). Consequently, the vectors ${\bf{g}}^{\prime, \prime\prime}$ come out} parallel (or antiparallel) to  ${\bf{n}}$. Thus, in this case, the energy flux also propagates along the same propagation direction of the electromagnetic wave, whatever the $\tilde{\alpha}'$, $\tilde{\alpha}''$ parameters values.

{For the general case in which $\mathbf{n}$ and $\mathbf{d}$ are not collinear or perpendicular, the energy flux is no longer parallel to the propagation direction of the electromagnetic wave.}

{\section{Bi-anisotropic case with antisymmetric parameters\label{bianisotropic_Asymm}}}

Now we analyze the case where the magnetoelectric parameters are described
by antisymmetric tensors, written as
\begin{subequations}
\label{constitutive-relations-anisymmetric-1}
\begin{align}
\alpha _{ij}& =   {\varepsilon _{ijk}}      a_{k},  \label{anti1} \\
\beta _{kn}& =   {\varepsilon _{knr}}    b_{r},  \label{anti2}
\end{align}
\end{subequations}
where $\mathbf{a}=(a_{x},a_{y},a_{z})$ and $\mathbf{b}=(b_{x},b_{y},b_{z})$
are fixed and, in principle, complex 3-vectors, which induce preferred
direction in the system, {while $\varepsilon _{ijk}$ } represents the usual
Levi-Civita symbol in three dimensions. In order to {satisfy}
Eq.~\eqref{relation-parameter3}, the following condition should hold:
\begin{equation}
\mathbf{b}^{*}=\mathbf{a}.  \label{a=b*}
\end{equation}
In the case the vectors $\mathbf{a}$ and $\mathbf{b}$ are real, this condition reduces merely to $\mathbf{b}=\mathbf{a}$.

{ Under the validity of relations (\ref{anti1}) and (\ref{anti2}),} the electric displacement field and the magnetic field are given by
\begin{align}
\mathbf{D}& =\epsilon \mathbf{E}+\mathbf{a}\times \mathbf{B},
\label{constitutive-anti1} \\
\mathbf{H}& =\frac{1}{\mu }\mathbf{B}+\mathbf{b}\times \mathbf{E}.
\label{constitutive-anti2}
\end{align}%
Some analog antisymmetric constitutive relations have found application in
the description of electron gas systems \cite{Carvalho} and in the investigation of electromagnetic propagation in time-dependent media with an antisymmetric magnetoelectric coupling and an isotropic time-dependent permittivity \cite{Lin}.

In the momentum space, Eq.\eqref{constitutive-anti1} provides
\begin{align}
\mathbf{D} &=\epsilon \mathbf{E}+\frac{1}{\omega }\mathbf{a}\times \left(
\mathbf{\mathbf{k}\times \mathbf{E}}\right) , \\
\mathbf{D}&=\left( \epsilon -\frac{\mathbf{a}\cdot \mathbf{\mathbf{k}}}{%
\omega }\right) \mathbf{\mathbf{E}}+\frac{\mathbf{a}\cdot \mathbf{\mathbf{E}}%
}{\omega }\mathbf{\mathbf{k}},
\end{align}
where $\mathbf{a}\times \left( \mathbf{\mathbf{k}\times \mathbf{E}}\right)
=\left( \mathbf{a}\cdot \mathbf{\mathbf{E}}\right) \mathbf{\mathbf{k-}}%
\left( \mathbf{a}\cdot \mathbf{\mathbf{k}}\right) \mathbf{\mathbf{E}}$. From
the relation $\mathbf{k}\cdot \mathbf{D}=0$, for $\mathbf{k}=\omega \mathbf{%
n}$, we obtain
\begin{equation}
\left[ \epsilon \mathbf{n}+\mathbf{n\times }\left( \mathbf{a\times n}\right) %
\right] \cdot \mathbf{\mathbf{E}}=0.  \label{condition1A}
\end{equation}

In general, the propagating modes are no longer transversal.
Yet, when $\mathbf{a}$ and $\mathbf{n}$ are parallel vectors, Eq. (\ref%
{condition1A}) becomes
\begin{equation}
\epsilon \mathbf{n}\cdot \mathbf{\mathbf{E}}=0,  \label{a_e_n_paralelos}
\end{equation}%
recovering a transversal electric field.

Replacing Eqs.~\eqref{anti1} and \eqref{anti2} in Eq.~\eqref{ex7}, one obtains the following extended electric permittivity tensor:
\begin{equation}
\label{anti3}
\bar{\epsilon}_{ij}=\left( \epsilon -\frac{%
	\mathbf{a}\cdot \mathbf{k}}{\omega }-\frac{\mathbf{b}\cdot \mathbf{k}}{%
	\omega }\right) \delta _{ij}+\frac{b_{i}k_{j}}{\omega }+\frac{a_{j}k_{i}}{%
	\omega },
\end{equation}
where there appear direction dependent terms: $(\mathbf{a}\cdot \mathbf{%
	k})$, $(\mathbf{b}\cdot \mathbf{k})$, $a_{i}k_{j}$, $a_{j}k_{i}$. Now the matrix
$M_{ij}$ is written as
\begin{align}
{M}&= \mathcal{N} +\mu [({\bf{a}}+{\bf{b}})\cdot {\bf{n}}] \mathbb{1}_{3\times 3} -\mu \left( \mathcal{A} +\mathcal{B} \right) , 	\label{anti4}
\end{align}
where $\mathcal{N}$ is given by Eq.\eqref{Nij1}, and
\begin{subequations}
\begin{align}
\mathcal{A} &= \mathrm{diag}\left(\frac{}{}(b_{1}+a_{1})n_{1}, (b_{2}+a_{2})n_{2}, (b_{3}+a_{3})n_{3}\right) , \\
\mathcal{B} &=\begin{pmatrix}
0 & b_{1}n_{2}+a_{2}n_{1} & b_{1}n_{3} + a_{3} n_{1} \\
b_{2}n_{1}+a_{1}n_{2} & 0 & b_{2}n_{3}+a_{3} n_{2} \\
b_{3}n_{1}+a_{1}n_{3} & b_{3}n_{2}+a_{2} n_{3} & 0
\end{pmatrix} .
\end{align}
\end{subequations}

For $\mathrm{det}[M_{ij}]=0$, the following
dispersion equation {is attained}:
\begin{align}
 0&=\left[ n^{2}-\mu \epsilon +\mu (\mathbf{c}\cdot
\mathbf{n})\right]\left\{\left[ n^{2}-\mu\epsilon +\mu(\mathbf{c}\cdot \mathbf{n})\right]\frac{}{}\right.\nonumber\\
&\left.\hspace{1cm}+\frac{\mu}{\epsilon} \left[ (\mathbf{a}\cdot \mathbf{b})n^{2}-(\mathbf{a}%
\cdot \mathbf{n})(\mathbf{b}\cdot \mathbf{n})\right]\right\} ,\label{anti5b}
\end{align}
which can be cast in the form:
\begin{equation}
\left[ \frac{{}}{{}}n^{2}-\mu  \epsilon  +\mu (\mathbf{c}%
\cdot \mathbf{n})\right] \left[ \mathbf{n}^{T}\mathbb{\tilde{Q}}\mathbf{n}%
+\mu (\mathbf{c}\cdot \mathbf{n})-\mu \epsilon \right]
=0,
\end{equation}%
where $\mathbf{c}=\mathbf{a}+\mathbf{b}$ is always a real vector [due to Eq. (\ref{a=b*})] and {$\mathbb{Q}$ is a complex $\left( 3\times 3\right) $ self-adjoint matrix} defined by
\begin{equation}
\mathbb{Q}=1+\frac{\mu \left( \mathbf{a}\cdot \mathbf{b}\right) }{\epsilon}-\frac{\mu \left( \mathbf{ab}^{T}+\mathbf{ba}^{T}\right) }{2\epsilon}.
\end{equation}

The first dispersion relation, expressed as
\begin{equation}
\left( \mathbf{n}+\frac{\mu {\mathbf{c}}}{2}\right) ^{2}=\mu \epsilon +\frac{%
\mu ^{2}{\mathbf{c}}^{2}}{4},
\end{equation}%
describes a sphere centered in $\mathbf{n}_{0}=-\mu {\mathbf{c}}/2$ with radius $\sqrt{\mu \epsilon +\mu ^{2}{\mathbf{c}}^{2}/4}$. The second dispersion relation is
\begin{equation}
\mathbf{n}^{T}\mathbb{Q}\mathbf{n}+\mu (\mathbf{a}+\mathbf{b})\cdot \mathbf{n%
}=\mu \epsilon.
\end{equation}%
The eigenvalues of $\mathbb{Q}$ determine the surface described by the
dispersion relation. If all eigenvalues are positive, the
surface becomes an ellipsoid whose center is not at the origin [if $(\mathbf{%
a}+\mathbf{b)}\cdot \mathbf{n}=0$, it is centered at the origin], with the
principal axes oriented along the respective eigenvectors. %
When there are at least two distinct eigenvalues, the
medium produces birefringence.

{\subsection{Propagation properties of magnetoelectric parameters in dielectrics}}

\begin{figure}[H]
\begin{centering}
\includegraphics[scale=.52]{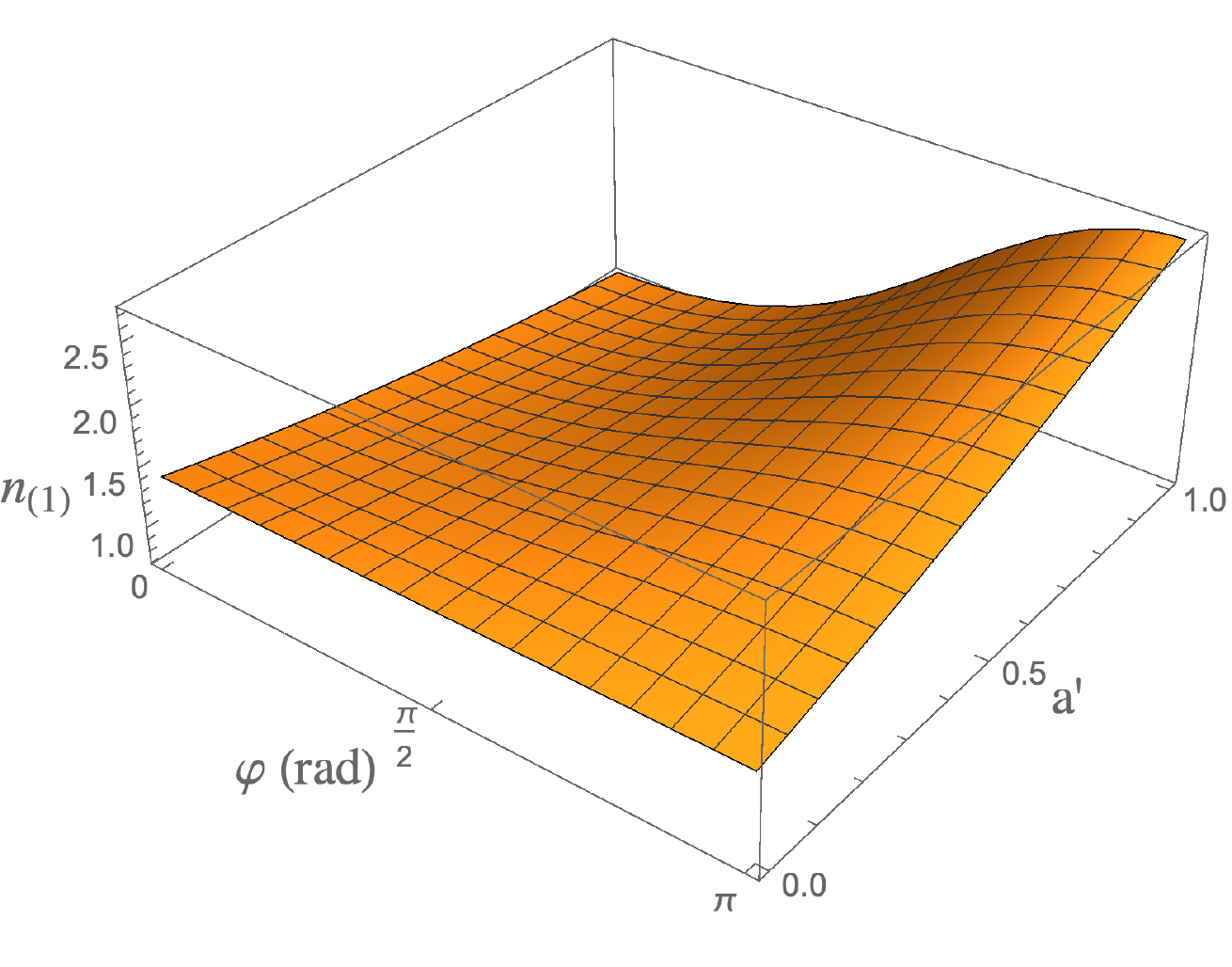}
\par\end{centering}
\caption{\label{plot-indice-refracao-1-caso-anti-simetrico}Refractive index $n_{(1)}$ of Eq.~\eqref{anti15A} in terms of $\varphi$ and $a'$. Here we have set $\mu=1$ and $\epsilon=2$. The parameters $\epsilon$, $\mu$, and $a'$ are dimensionless.}
\end{figure}

\begin{figure}[H]
\begin{centering}
\includegraphics[scale=.52]{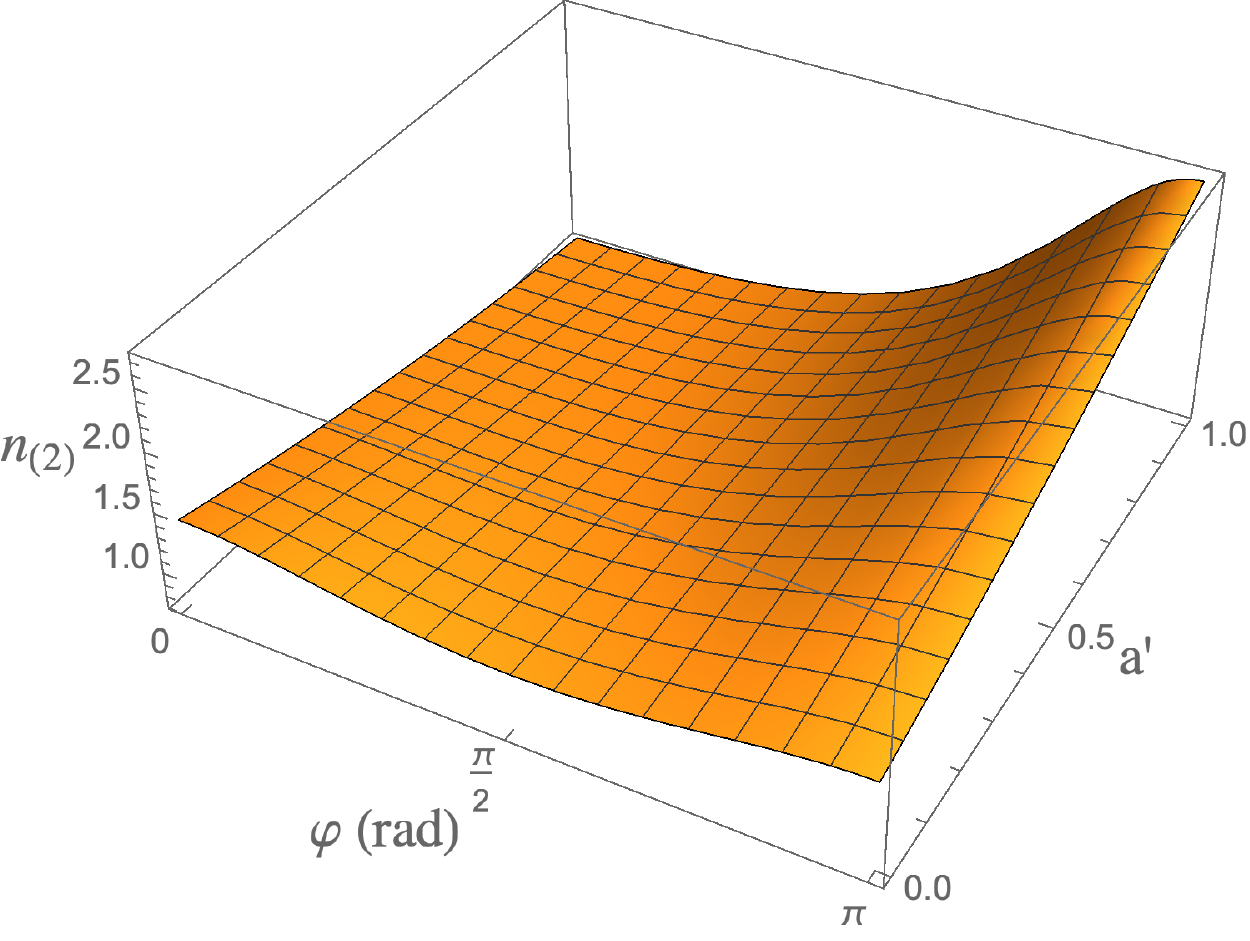}
\par\end{centering}
\caption{\label{plot-indice-refracao-2-caso-anti-simetrico}Refractive index $n_{(2)}$ of Eq.~\eqref{anti15B} in terms of $\varphi$ and $a'$. We have used $\mu=1$, $\epsilon=2$ and $a''=1$. {The parameters $\epsilon$, $\mu$, and $a'$ are dimensionless.}}
\end{figure}

In order to investigate the electromagnetic propagation in a {dielectric medium governed} by the constitutive relations (\ref{constitutive-anti1}) and (\ref{constitutive-anti2}), we suppose that the magnetoelectric parameters are constrained by relation \eqref{a=b*}. In the case the vectors $\mathbf{a}$ and
$\mathbf{b}$ have a complex piece, that is
\begin{equation}
\mathbf{a}=\mathbf{a}^{\prime }+\mathrm{i} \mathbf{a}^{\prime \prime },
\quad \mathbf{b}=\mathbf{b}^{\prime }+\mathrm{i} \mathbf{b}^{\prime \prime },
\label{complexVectors}
\end{equation}
one writes $\mathbf{a}^{\prime }=\mathbf{b}^{\prime }$ and $\mathbf{a}^{\prime
	\prime }=-\mathbf{b}^{\prime \prime }$, as a consequence of Eq.~\eqref{a=b*}. In this case, we have
\begin{align}
(\mathbf{a}+\mathbf{b}) \cdot \mathbf{n}=2(\mathbf{a}^{\prime }\cdot \mathbf{%
	n}),  \label{real1A} \\
(\mathbf{a}\cdot \mathbf{n}) (\mathbf{b}\cdot \mathbf{n})=(\mathbf{{%
		a^{\prime }}}\cdot \mathbf{n})^{2}+(\mathbf{{a^{\prime \prime }}}\cdot
\mathbf{n})^{2},  \label{real1B} \\
\mathbf{a}\cdot \mathbf{b}={a^{\prime }}^2+{a^{\prime\prime }}^{2}=|a|^{2},
\end{align}
for a real $\mathbf{n}$ vector. We also suppose that the 3-vector $\mathbf{a}$ fulfills $\mathbf{a}^{\prime } \cdot
\mathbf{n}= a^{\prime }n\cos\varphi$, $\mathbf{a}^{\prime\prime } \cdot
\mathbf{n}= a^{\prime\prime }n\cos\varphi$, so that the involved relation (\ref{anti5b})
provides,
\begin{subequations}
\label{anti5c-1-1}
\begin{align}
\left[n^{2}-\mu\epsilon + 2\mu a^{\prime } n \cos\varphi\right]=0 \label{anti5c} \\
\left[n^{2} \left(\epsilon+\mu |a|^{2}\sin^{2}\varphi \right) + 2 \mu
\epsilon a^{\prime } n \cos\varphi - \mu \epsilon^{2} \right]=0 , \label{anti5c1}
\end{align}
\end{subequations}
from which the following (positive) indices are achieved:
\begin{align}
n_{(1)} &= \mu \sqrt{{a^{\prime }}^2 \cos^{2}\varphi+\epsilon/\mu}-\mu a^{\prime } \cos\varphi ,
\label{anti15A} \\
n_{(2)}&=\frac{1}{r} \left( \sqrt{\mu\epsilon+\mu^{2}{a^{\prime}}^2+\mu^{2}{a^{\prime\prime}}^2 \sin^{2}\varphi }  -\mu a^{\prime}
\cos\varphi \right),  \label{anti15B}
\end{align}
where
\begin{align}  \label{antisymmetric-new-1}
r&= 1 +\frac{\mu}{\epsilon} |a|^{2} \sin^{2} \varphi .
\end{align}
Above, we have retained only the roots
corresponding to positive refractive indices, since we are not addressing
metamaterials. {The general behavior of $n_{(1, 2)}$ is illustrated in Figs. \ref{plot-indice-refracao-1-caso-anti-simetrico} and \ref{plot-indice-refracao-2-caso-anti-simetrico} in terms of $\varphi \in [0, \pi]$ and {the dimensionless parameter} $a' \in [0, 1]$. }

The anisotropy effect is described by the angular dependence on $\varphi$, {not exactly equal for $n_{(1)}$ and $n_{(2)}$. In fact, note that by setting $a' \mapsto 0$, one obtains $n_{1} \mapsto \sqrt{\mu\epsilon}$, a constant value which corresponds to the straight border line of Fig. \ref{plot-indice-refracao-1-caso-anti-simetrico} (for $a' = 0$). Under such a limit, $n_{2} \mapsto \mu \epsilon \left( \mu\epsilon + \mu^{2} a''^{2} \sin^{2}\varphi \right)^{-1/2}$, which is represented by the sinusoidal border line at $a'=0$ in Fig. \ref{plot-indice-refracao-2-caso-anti-simetrico}. Furthermore, the border behavior at $a'=1$ is also distinct for the two plots.} Both $n_{(1)}$ and $n_{(2)}$ have maximal values at $\varphi=\pi$.

{It is important to mention that both Eq.~\eqref{anti5b} as the refractive indices \eqref{anti15A} and \eqref{anti15B} hold for any propagation direction in relation to the vector $\mathbf{a}$, parametrized by the angle $\varphi$. Now, we investigate propagation for some special angles} between $\mathbf{a}$ and $\mathbf{n}$.

\bigskip

\subsubsection{Particular case 1: $\mathbf{a}$-orthogonal configuration. \label{case1ants}} 

We begin bychoosing a scenario where the vector $\mathbf{a}$ is orthogonal to the
propagation axis, that is, $\mathbf{a}\cdot \mathbf{n}=0$ (or $\varphi=\pi/2$%
). Thus, the dispersion relation (\ref{anti5c-1-1}) provides two solutions for
the refractive index,
\begin{align}
n_{(1)}&=\sqrt{\mu \epsilon},  \label{anti9A} \\
n_{(2)}&=\frac{\sqrt{\mu \epsilon}}{\sqrt{1+(\mu/\epsilon){|a|}^2}}.
\label{anti9B}
\end{align}
The two refractive indices are real. Therefore, there occurs birefringence but not absorption. To obtain the
propagating modes, we take the vector $\mathbf{n}$ along the $z$-axis, that is, $\mathbf{n} =(0,0,n_{3})$.
Orthogonal to it, we set $\mathbf{{a^{\prime}}}=(a_{1},a_{2},0)$, $\mathbf{{a^{\prime\prime}}}=(a_{1},a_{2},0)$.
This choice leads to a very simple expression for the matrix of Eq.~\eqref{anti4},
that is,
\begin{equation}
{M} =%
\begin{pmatrix}
n_{3}^{2}-\mu \epsilon & 0 & -\mu a_{1}^{*} n_{3} \\
0 & n_{3}^{2}-\mu \epsilon & -\mu {a_{2}^{*}} n_{3} \\
-\mu {a_{1}} n_{3} & -\mu {a_{2}}  n_{3} & -\mu \epsilon%
\end{pmatrix}
.  \label{anti110}
\end{equation}
Replacing index (\ref{anti9A}) in matrix
(\ref{anti110}), the condition $M_{ij}E^{j}=0$ provides as a solution a transversal mode,
\begin{equation}
\mathbf{E}_{(1)}=\frac{1}{|\mathbf{a}|}
\begin{pmatrix}
{a_{2}} \\
-{a_{1}} \\
0%
\end{pmatrix}
,  \label{anti18A}
\end{equation}
with $|\mathbf{a}|=\sqrt{|{a_{1}}|^2+|{a_{2}}|^2}$.
Now, replacing relation (\ref{anti9B}) in matrix (\ref{anti110}), one has
\begin{equation}
\begin{pmatrix}
-\mu \epsilon f_{a} & 0 & -\mu {a_{1}^{*}}n_{3} \\
0 & -\mu \epsilon f_{a} & -\mu {a_{2}^{*}}n_{3} \\
-\mu {a_{1}}n_{3} & -\mu {a_{2}}n_{3} & -\mu \epsilon%
\end{pmatrix}
\begin{pmatrix}
E_{x} \\
E_{y} \\
E_{z}%
\end{pmatrix}
=0,  \label{anti112}
\end{equation}
whose solution is
\begin{equation}
\mathbf{E}_{(2)}=\frac{1}{|a|\sqrt{1+f_{a}}}
\begin{pmatrix}
{a_{1}^{*}} \\
{a_{2}^{*}} \\
-|a|\sqrt{f_{a}}%
\end{pmatrix}
,  \label{anti18B}
\end{equation}
with
\begin{align}  \label{antisymmetric-new-2}
f_{a} &= \frac{(\mu/\epsilon)|a|^2}{1+(\mu/\epsilon){|a|^2}} .
	\end{align}
We note that Eq.\eqref{anti18B} represents a mixed mode,
endowed with a longitudinal component. In this case, it is
not possible to find a pure transversal mode for the field $\mathbf{E}_{(2)}$.
The transversal mode $\mathbf{E}_{(1)}$ and the transversal sector of the mixed mode $\mathbf{E}_{(2)}$ could exhibit linear polarization, or circular or elliptical polarization.  Indeed, in principle one writes the $\mathbf{{a}}$-vector  components as $a_1=(a_{1}^{\prime}+\mathrm{i}a_{1}^{\prime\prime})$, $a_2=(a_{2}^{\prime}+\mathrm{i}a_{2}^{\prime\prime})$. One notices that for either $a_{1}''=a_{2}''=0$ or $a_{1}'=a_{2}'=0$, Eq.~\eqref{anti18A} and the transversal part of Eq.~\eqref{anti18B} yield linearly polarized modes. On the other hand, for either $a_{1}'=a_{2}''=0$ or $a_{1}''=a_{2}'=0$, that is, for $a_1=a_{1}^{\prime}$, $a_2=\mathrm{i}a_{2}^{\prime\prime}$, or $a_1=\mathrm{i}a_{1}^{\prime\prime}$, $a_2=a_{2}^{\prime}$, the polarization is elliptical. Circularly polarized modes only occur when either $a_{1}'=a_{2}''=0$ and $a_{2}'=a_{1}''$ or $a_{1}''=a_{2}'=0$ and $a_{1}'=a_{2}''$.

After finding the refractive indices (\ref{anti9A}) and (\ref{anti9B}), and the
corresponding modes (\ref{anti18A}) and (\ref{anti18B}), we need to discuss the
physical effects on wave propagation. In the case the associated modes are
linearly or elliptically polarized, the implied birefringence
is expressed in terms of the phase shift \eqref{phase-shift-0}, namely,

\begin{equation}
\Delta=\frac{2\pi}{\lambda_{0}} l [n_{(1)}-n_{(2)}]\,,  \label{phase-shift1}
\end{equation}
where $n_{1}$ and $n_{2}$
are the refractive indices (\ref{anti9A}) and (\ref{anti9B}), respectively. The phase shift per unit length is

\begin{equation}
\frac{\Delta}{l}=\frac{2\pi}{\lambda_{0}} \sqrt{\mu \epsilon} \left[1-\frac{1}{\sqrt{1+(\mu/\epsilon)|a|^2}}\right]\,,
		\label{phase-shift2}
		\end{equation}
	
	which, for $(\mu/\epsilon)|a|^2 \ll1$, simplifies as
	\begin{equation}
\frac{\Delta}{l}=\frac{\pi\mu |a|^2}{\lambda_{0}}.
\label{phase-shift2B}
\end{equation}

\subsubsection{Particular case 2: $\mathbf{a}$-longitudinal configuration.}

Let us now consider the case where the vectors $\mathbf{a}$ and $\mathbf{n}$
point along the
same direction, $\mathbf{a}
\cdot \mathbf{n}= a n $, for which Eq.~\eqref{anti5b} is written as
\begin{align}  \label{anti5d}
\left( n^{2} +2\mu a^{\prime} n -\mu\epsilon \right)^{2} =0 ,
\end{align}
which involves the square of a quadratic polynomial in $n$. Thus, there is a
doubly degenerate refractive index, namely,
\begin{align}  \label{antysimmetric-new-3}
n=\sqrt{\mu\epsilon+\mu^{2}{a^{\prime}}^{2}}-\mu a^{\prime}.
\end{align}
The latter corresponds exactly to the solutions of Eqs.~\eqref{anti15A} and %
\eqref{anti15B} for $\varphi=0$, as expected.

To obtain the propagating modes, we take the vector $\mathbf{n}$
along the $z$-axis, that is, $\mathbf{n} =(0,0,n_{3})$, in such a way that $\mathbf{a}%
=\mathbf{b^{*}}=(0,0,a_{3}^{\prime}+\mathrm{i}a_{3}^{\prime\prime})$. It leads to a simple form of the matrix of %
Eq.\eqref{anti4}, that is,
\begin{equation}
{M} =
\begin{pmatrix}
n_{3}^{2}-\mu \epsilon +A & 0 & 0 \\
0 & n_{3}^{2}-\mu \epsilon +A & 0 \\
0 & 0 & -\mu \epsilon%
\end{pmatrix}%
,  \label{anti111}
\end{equation}
where $A=2\mu na_{3}^{\prime}$. With index (\ref{antysimmetric-new-3}),
matrix (\ref{anti111}) reads simply as
\begin{equation}
{M}=
\begin{pmatrix}
0 & 0 & 0 \\
0 & 0 & 0 \\
0 & 0 & -\mu \epsilon%
\end{pmatrix}%
,  \label{anti112a}
\end{equation}
and provides generic transversal modes,
\begin{equation}
\mathbf{E}=
\begin{pmatrix}
E_{x} \\
E_{y} \\
0%
\end{pmatrix}
,  \label{anti19B}
\end{equation}
with arbitrary $E_{x}$ and $E_{y}$. Note this transversality occurs in accordance with Eq. (\ref{a_e_n_paralelos}). The solution (\ref{anti19B}) may represent a
linear, circular, or elliptic polarization mode, depending on the nature and relation
between $E_{x}$ and $E_{y}$. As this relation is not supplied by the
features examined so far, we conclude that the $\mathbf{a}$-longitudinal
configuration allows any polarization, in principle. Furthermore, only one refractive index expression was achieved, with no signal of anisotropy. Thus, we conclude that the ${\bf{a}}$ direction defines the optical axis of the medium.

 It is worthwhile to observe that the complex vectors $\mathbf{a}$ and
$\mathbf{b}$, given in Eq.~\eqref{complexVectors}, yield real refractive indices as it happens in the case of real vectors.
For the $\mathbf{a}$-orthogonal configuration, the complex vectors of Eq.~\eqref{complexVectors} may provide circular or elliptical polarization (in its transversal sector), besides the linear one.

{\subsection{Group velocity, phase velocity, and Poynting vector}}

{From Eq. \eqref{anti5b}, we can write down two dispersion equations,
\begin{align}
0&=\omega^{2} - 2\frac{\omega}{\epsilon} ({\bf{a}}' \cdot {\bf{k}}) - \frac{ {\bf{k}}^{2}}{\mu\epsilon} , \label{group-velocity-antisymmetric-1}\\
0 &=\omega^{2} -2 \frac{\omega}{\epsilon} ({\bf{a}}' \cdot {\bf{k}}) - \left( \frac{1}{\mu\epsilon} + \frac{ |a|^{2}}{\epsilon^{2}} \right) {\bf{k}}^{2} + \frac{ ({\bf{a}}' \cdot {\bf{k}})^{2}}{\epsilon^{2}} +\nonumber \\
&\phantom{=}+ \frac{ ( {\bf{a}}'' \cdot {\bf{k}})^{2}}{\epsilon^{2}}, \label{group-velocity-antisymmetric-2}
\end{align}
which provide, respectively, the following solutions for $\omega$ (with $\omega>0$):
\begin{align}
\omega_{(1)}&= \sqrt{\frac{ {\bf{k}}^{2}}{\mu\epsilon} + \frac{ ({\bf{a}}' \cdot {\bf{k}})^{2}}{\epsilon^{2}}} + \frac{ ({\bf{a}}' \cdot {\bf{k}})}{\epsilon} , \label{group-velocity-antisymmetric-3} \\[0.2cm]
\omega_{(2)} &= \sqrt{ \left(\frac{1}{\mu\epsilon} + \frac{ |a|^{2}}{\epsilon^{2}} \right) {\bf{k}}^{2} - \frac{ ({\bf{a}}'' \cdot {\bf{k}})^{2}}{\epsilon^{2}} } + \frac{ ({\bf{a}}' \cdot {\bf{k}})}{\epsilon}. \label{group-velocity-antisymmetric-4}
\end{align}
}

{
The phase and group velocities are given by
\begin{align}
v_{\mathrm{ph}(1)} &= \sqrt{ \frac{1}{\mu\epsilon} + \frac{ ({\bf{a}}' \cdot \hat{\bf{k}})^{2}}{\epsilon^{2}}} + \frac{ ({\bf{a}}' \cdot \hat{\bf{k}})}{\epsilon}, \label{group-velocity-antisymmetric-5} \\
v_{\mathrm{ph}(2)} &= \sqrt{ \frac{1}{\mu\epsilon} + \frac{ |a|^{2}}{\epsilon^{2}} - \frac{ ({\bf{a}}'' \cdot \hat{\bf{k}})^{2}}{\epsilon^{2}} } + \frac{ ({\bf{a}}' \cdot \hat{\bf{k}})}{\epsilon}, \label{group-velocity-antisymmetric-6} \\
{\bf{v}}_{\mathrm{g}(1)} &= f_{a'} \left[ \frac{ {\bf{k}}}{\mu\epsilon} + \frac{ ({\bf{a}}' \cdot {\bf{k}})}{\epsilon^{2}} {\bf{a}}' \right] + \frac{ {\bf{a}}'}{\epsilon} , \label{group-velocity-antisymmetric-7} \\
{\bf{v}}_{\mathrm{g}(2)} &= f_{a''} \left[ \left(\frac{1}{\mu\epsilon} + \frac{|a|^{2}}{\epsilon^{2}} \right) {\bf{k}} - \frac{ ({\bf{a}}'' \cdot {\bf{k}})}{\epsilon^{2}} {\bf{a}}'' \right] + \frac{ {\bf{a}}'}{\epsilon} , \label{group-velocity-antisymmetric-8}
\end{align}
with
\begin{align}
f_{a'} &= \left[ \frac{ {\bf{k}}^{2}}{\mu\epsilon} + \frac{ ({\bf{a}}' \cdot {\bf{k}})^{2}}{\epsilon^{2}} \right] ^{-1/2} , \label{group-velocity-antisymmetric-9} \\
f_{a''} &=  \left[ \left( \frac{1}{\mu\epsilon} + \frac{ |a|^{2}}{\epsilon^{2}} \right) {\bf{k}}^{2} - \frac{ ({\bf{a}}'' \cdot {\bf{k}})^{2}}{\epsilon^{2}} \right]^{-1/2} . \label{group-velocity-antisymmetric-10}
\end{align}
}

{Relations \eqref{group-velocity-antisymmetric-7} and \eqref{group-velocity-antisymmetric-8} reveal that the group velocity is in general not parallel to ${\bf{n}}$.}
{Starting from Eqs. (\ref{group-velocity-antisymmetric-3}) and (\ref{group-velocity-antisymmetric-4}), we now particularize the phase and group velocities for propagation orthogonal and longitudinal to  ${\bf{a}}'$.}

\begin{figure}[H]
\begin{centering}
\includegraphics[scale=.6]{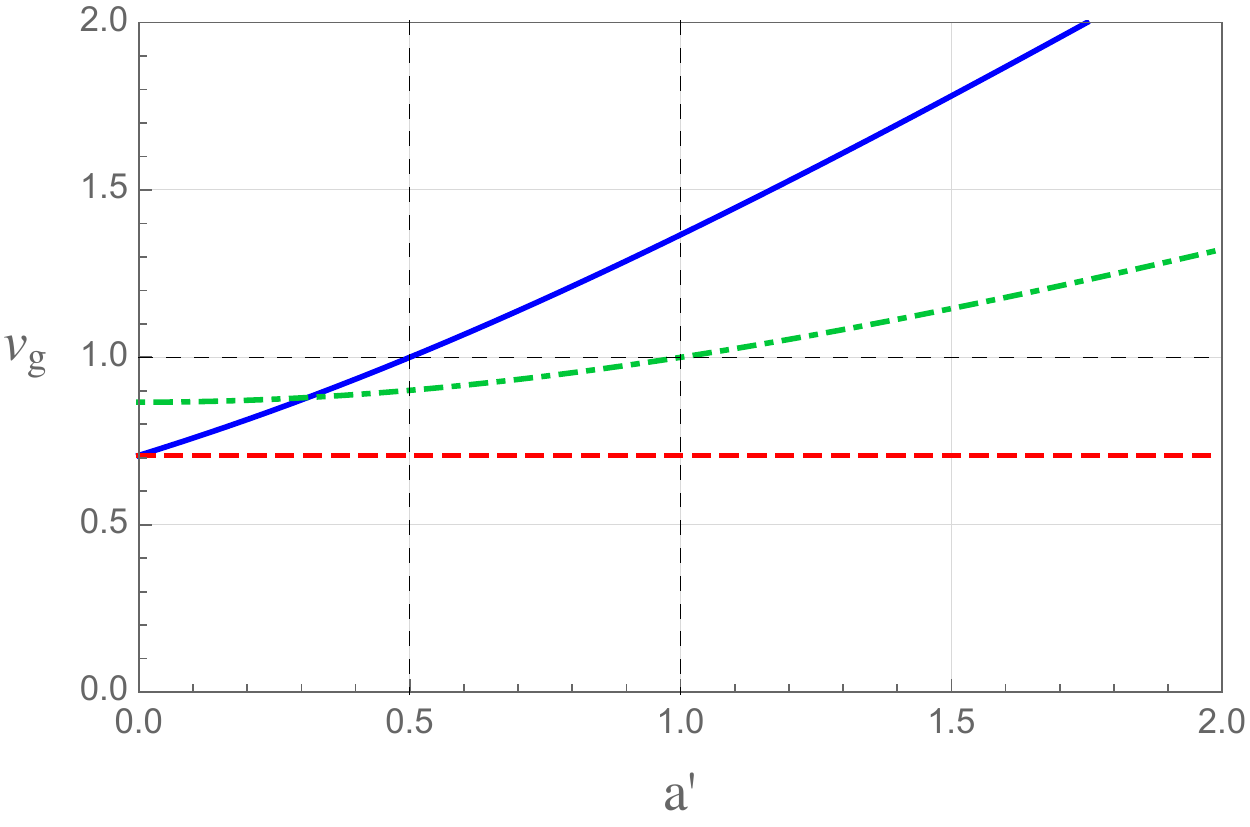}
\par\end{centering}
\caption{\label{plot-group-velocities-antisymmetric-longitudinal-and-ortogonal} Group velocities for ${\bf{a}}$-longitudinal and ${\bf{a}}$-orthogonal cases. The blue solid line indicates $v_{\mathrm{g}(1, 2)}^{\mathrm{long.}}$ of Eqs.~\eqref{group-velocity-antisymmetric-13} and \eqref{group-velocity-antisymmetric-14}. The red curve illustrates $v_{\mathrm{g}(1)}^{\mathrm{orth.}}$ of Eq.~\eqref{group-velocity-antisymmetric-11}, and the green line represents $v_{\mathrm{g}(2)}^{\mathrm{orth.}}$ of Eq.~\eqref{group-velocity-antisymmetric-12}. The gray dashed vertical lines indicate the values of $a' \in \{1/2, 1\}$ above which the group velocities become greater than 1, in agreement with Eqs. (\ref{group-velocity-antisymmetric-15}), and (\ref{group-velocity-antisymmetric-17}), respectively. Here we have used $\mu=1$, $\epsilon=2$ and $a''=1$.}
\end{figure}

\begin{itemize}
{\item For the ${\bf{a}}$-orthogonal case where ${\bf{a}}'\cdot {\bf{k}}={\bf{a}}'' \cdot {\bf{k}}=0$, one finds
\begin{align}
v_{\mathrm{ph}(1)}^{\mathrm{orth}} &=\sqrt{\frac{1}{\mu\epsilon}}, \quad  \mathbf{v}_{\mathrm{g}(1)}^{\mathrm{orth}} = \hat{\bf{k}} \sqrt{\frac{1}{\mu\epsilon}} , \label{group-velocity-antisymmetric-11} \\[0.2cm]
v_{\mathrm{ph}(2)}^{\mathrm{orth}}  &=   \sqrt{\frac{1}{\mu\epsilon} + \frac{ |a|^{2}}{\epsilon^{2}} },  \quad {\bf{v}}_{\mathrm{g}(2)}^{\mathrm{orth}} = \hat{\bf{k}} \sqrt{\frac{1}{\mu\epsilon} + \frac{|a|^{2}}{\epsilon^{2}}}.
 \label{group-velocity-antisymmetric-12}
\end{align}}
{\item For the ${\bf{a}}$-longitudinal scenario ${\bf{a}}^{\prime, \prime\prime} \cdot {\bf{k}} = a^{\prime, \prime\prime} k $, we obtain
\begin{align}
v_{\mathrm{ph}(1)}^{\mathrm{long}} &= v_{\mathrm{g}(1)}^{\mathrm{long}} =\sqrt{ \frac{1}{\mu\epsilon} + \frac{a'^{2}}{\epsilon^{2}}} + \frac{a'}{\epsilon}, \label{group-velocity-antisymmetric-13} \\
v_{\mathrm{ph}(2)}^{\mathrm{long}} &= v_{\mathrm{g}(2)}^{\mathrm{long}} = \sqrt{\frac{1}{\mu\epsilon} + \frac{a'^{2}}{\epsilon^{2}
}} + \frac{a'}{\epsilon}, \label{group-velocity-antisymmetric-14}
\end{align}
with
\begin{equation}
\quad {\bf{v}}_{\mathrm{g}(1)}^{\mathrm{orth}} = v_{\mathrm{g}(1)}^{\mathrm{orth}}\hat{\bf{k}},	\quad {\bf{v}}_{\mathrm{g}(2)}^{\mathrm{orth}} = v_{\mathrm{g}(2)}^{\mathrm{orth}}\hat{\bf{k}}.
\end{equation}}
\end{itemize}
{In Fig.~\ref{plot-group-velocities-antisymmetric-longitudinal-and-ortogonal} we plot the group velocities $v_{\mathrm{g}(1, 2)}$ of both ${\bf{a}}$-longitudinal and ${\bf{a}}$-orthogonal cases in terms of the parameter $a'$.}

{We notice the group velocities can be greater than 1 depending on the values of $a'$.
\begin{itemize}
\item For the ${\bf{a}}$-longitudinal case, in order to ensure $v_{\mathrm{g}}^{\mathrm{long}}<1$, the following must occur:
 \begin{align}
 a'< \frac{\mu\epsilon-1}{2\mu} . \label{group-velocity-antisymmetric-15}
 \end{align}
 \item For the ${\bf{a}}$-orthogonal configuration, one finds that $v_{\mathrm{g}(2)}^{\mathrm{orth}}<1$  only when
 \begin{align}
 |a|< \sqrt{\frac{\epsilon}{\mu} (\mu\epsilon-1)} . \label{group-velocity-antisymmetric-17}
 \end{align}
\end{itemize}
These results are important to constrain the magnetoelectric parameters in ranges suitable to ensure group velocity smaller than 1.}

{With the constitutive relation (\ref{constitutive-anti2}), the Poynting vector is
\begin{align}
{\bf{S}} &= \frac{1}{2\mu} ({\bf{n}}+\mu {\bf{a}}) |{\bf{E}}|^{2} - \frac{1}{2\mu} ({\bf{n}}\cdot {\bf{E}} ) {\bf{E}}^{*} - \frac{1}{2} ({\bf{a}}\cdot {\bf{E}}) {\bf{E}}^{*}. \label{poynting-vector-antisymmetric-1}
\end{align}}

{
By considering ${\bf{a}}= {\bf{a}}' + \mathrm{i} {\bf{a}}''$ and ${\bf{E}}={\bf{E}}' +\mathrm{i} {\bf{E}}''$, we firstly analyze the case $\bf{a}', \bf{a}''$ parallel to $\mathbf{n}$. From Eq. (\ref{condition1A}) we obtain  $\mathbf{n}\cdot{\bf{E}}=0$, i.e., $\mathbf{n}\cdot\mathbf{E}'=0=\mathbf{n} \cdot\mathbf{E}''$ implying $\mathbf{a}'\cdot\mathbf{E}=0=\mathbf{a}'' \cdot\mathbf{E}$ or $\mathbf{a}\cdot\mathbf{E}=0$. This way Eq. (\ref{poynting-vector-antisymmetric-1}) simplifies to
\begin{align}
{\bf{S}}^{\mathrm{long.}} &= \frac{1}{2\mu} ({\bf{n}}+\mu {\bf{a}}) |{\bf{E}}|^{2}, \label{poynting-vector-antisymmetric-1p}
\end{align}
whose time-averaged form becomes
\begin{align}
\left \langle {\bf{S}} \right \rangle^{\mathrm{long.}} &= \frac{1}{2\mu} ({\bf{n}}+\mu {\bf{a}'}) |{\bf{E}}|^{2}. \label{poynting-vector-antisymmetric-1q}
\end{align}
As ${\bf{a}'}\parallel {\bf{n}}$, the energy flux propagates along the wave propagation direction in this situation.}

{For a general case, the time-averaged Poynting vector is
\begin{align}
\left \langle {\bf{S}} \right \rangle &= \frac{1}{2\mu} ({\bf{n}}+ \mu {\bf{a}}' ) |{\bf{E}}|^{2} +  \nonumber \\[0.2cm]
&\phantom{=} +\frac{1}{2} (F_{1} {\bf{E}}' - F_{2} {\bf{E}}'')   \left[  ({\bf{a}}' \cdot {\bf{E}}') +({\bf{a}}'' \cdot {\bf{E}}'') \right] + \nonumber \\[0.2cm]
&\phantom{=} + \frac{1}{2} (F_{1} {\bf{E}}'' + F_{2} {\bf{E}}') \left[  ({\bf{a}}' \cdot {\bf{E}}'') - ({\bf{a}}'' \cdot {\bf{E}}') \right], \label{poynting-vector-antisymmetric-2}
\end{align}
where
\begin{eqnarray}
F_{1}=\frac{n^{2}}{\mu }\frac{\left[ \epsilon -\left( \mathbf{n\cdot a}
^{\prime }\right) \right] }{\Delta }-1, \quad
F_{2} =\frac{n^{2}}{\mu }\frac{\left( \mathbf{n\cdot a}^{\prime \prime
}\right) }{\Delta }, \label{poynting-vector-antisymmetric-4}
\end{eqnarray}
with
\begin{eqnarray}
\Delta  =\left[ \epsilon -\left( \mathbf{n\cdot a}^{\prime }\right) \right]
^{2}+\left( \mathbf{n\cdot a}^{\prime \prime }\right) ^{2}.\label{poynting-vector-antisymmetric-5}
\end{eqnarray}}

{
In particular, by considering  $\bf{a}', \bf{a}''$ orthogonal to $\mathbf{n}$, we obtain
\begin{align}
\left\langle {\bf{S}} \right\rangle^{\mathrm{orth.}} &= \frac{1}{2\mu} ({\bf{n}}+\mu {\bf{a}}' ) |{\bf{E}}|^{2} + \frac{(n^{2}-\mu\epsilon)}{2\mu\epsilon} \left[ \frac{}{} ({\bf{a}}' \cdot {\bf{E}}')  {\bf{E}}' + \right. \nonumber \\
&\left. \phantom{=} + ({\bf{a}}'' \cdot {\bf{E}}'')  {\bf{E}}'+({\bf{a}}' \cdot {\bf{E}}'')  {\bf{E}}''  - ({\bf{a}}'' \cdot {\bf{E}}')  {\bf{E}}''     \frac{}{}\right] , \label{poynting-vector-antisymmetric-6}
\end{align}
{which yields a propagation not aligned to $\bf{n}$. This result is consistent with the group velocities \eqref{group-velocity-antisymmetric-7} and \eqref{group-velocity-antisymmetric-8}, which contain a piece not belonging to the $\bf{n}$ axis.}}

{Finally, we note that for $\bf{a}', \bf{a}''$ nonparallel to ${\bf{n}}$, the energy flow does not propagate along the propagation axis because there is a contribution along the electric field direction.}

\section{\label{final-remarks}Final Remarks}

In this work, we have examined the propagation of electromagnetic waves in bi-isotropic and bi-anisotropic matter. As for the bi-anisotropic scenarios, we have taken isotropic electric permittivity and magnetic permeability tensors, while the magnetoelectric parameters were supposed as symmetric and antisymmetric anisotropic complex tensors. As an initial action, a modified permittivity tensor was written as part of the matrix { equation, which allowed us to obtain} the dispersion equations and refractive indices.

The symmetric magnetoelectric tensors were parametrized in terms of a single
3-vector, $\mathbf{d}$, and two complex scalars, $\tilde{\alpha}, \tilde{\beta%
}$. For propagation along the $z$-axis, $\mathbf{n} =(0,0,n)$, arbitrary
transversal propagating modes were obtained for the $\mathbf{d}$%
-longitudinal configuration. The $\mathbf{d}$ axis coincides with the optical axis of the medium. On the other hand, when the $\mathbf{d}$-vector is orthogonal to the propagation axis, the associated modes are also transversal, with polarization linear or elliptical.

The antisymmetric magnetoelectric tensors were parametrized in terms of two
3-vectors, $\mathbf{a}$ and $\mathbf{b}$, related by $\mathbf{b}^{*}=\mathbf{a}$. {We have considered complex vectors and obtained real and positive refractive indices} [see Eqs.~\eqref{anti15A} and \eqref{anti15B}]. For the  $\mathbf{a}$-longitudinal configuration, arbitrary transversal propagating modes, associated with one unique refractive index, were obtained for the propagation along the $z$ axis, $\mathbf{n} =(0,0,n)$. Therefore, in this case, the vector $\mathbf{a}$ determines the optical axis of the medium. For the $\mathbf{a}$-orthogonal configuration, there appears a transversal and a mixed mode, composed of a longitudinal and a transversal piece, associated with two refractive indices. The polarization may be linear, {elliptical, or circular.} The birefringence effect was evaluated in terms of the $\mathbf{a}$-vector magnitude.

In both symmetric and antisymmetric cases, the propagation along the $\mathbf{d}$ or $\mathbf{a}$ directions is isotropic, since they define the optical axis of the medium. On the other hand, the propagation orthogonal to the magnetoelectric vectors $\mathbf{d}$ or $\mathbf{a}$ provides a route of
phenomenological distinction between the symmetric [Eq.~(\ref{symmetric1})] or antisymmetric tensor [Eq.~(\ref{constitutive-relations-anisymmetric-1})], due to the observed difference between the associated propagating modes. See Eqs. \eqref{propagating-symmetric-6B-1} and \eqref{anti18B}.
{We have also evaluated the group velocities for the bi-isotropic and bi-anisotropic cases examined. In the bi-isotropic case, the phase and group velocities turned out equal. For the bi-anisotropic configurations, these velocities are different. The general group velocities present a component along the 3-vectors $\mathbf{d}$ or $\mathbf{a}$, being no longer necessarily parallel to $\mathbf{n}$ in these cases.}
{We have also carried out the Poynting vector, observing that the electromagnetic energy flux does not occur along the wave propagation direction for general configurations. } {The energy flux direction coincides with the $\mathbf{n}$ axis in the following situations: when $\mathbf{d}$ is parallel or orthogonal to $\mathbf{n}$ (in the symmetric case) and when $\mathbf{a}'$ and $\mathbf{a}'' $ are parallel to $\mathbf{n}$ (in the antisymmetric case).}

{As a final remark, we may try to state a parallel between our results and the physics of anisotropic media described by constitutive relations \eqref{SCR1B}, with permittivity and permeability described by general tensors $\epsilon_{ij}$ and $\mu_{ij}$. This is the case examined in Refs. \cite{Yakov, Markel}.  The general anisotropic dispersion relations of these references are different from our relations, which are based on isotropic permittivity and permeability, $\epsilon_{ij}=\epsilon\delta_{ij}$ and $\mu_{ij}=\mu\delta_{ij}$, and non-null magnetoelectric parameters, $\alpha_{ij}$ and $\beta_{ij}$. These basic distinctions, consequently, do not favor straightforward comparisons between such anisotropic systems.}

\subsection*{Acknowledgments}

The authors P.D.S.S., R.C., and M.M.F. express their gratitude to FAPEMA, CNPq, and CAPES (Brazilian research agencies) for invaluable financial support. R.C. acknowledges the support from the Grants No. CNPq/306724/2019-7, No. FAPEMA/Universal-01131/17, and No. FAPEMA/Universal-00812/19. M.M.F. is supported by No. FAPEMA/Universal/01187/18, No. CNPq/Produtividade 311220/2019-3, and No. CNPq/Universal/422527/2021-1. Furthermore, we are indebted to CAPES/Finance Code 001 and FAPEMA/POS-GRAD-02575/21.

\appendix

\section{\label{AppendixA} Vector evaluation for the electric field of the modes}
In this Appendix, we present an alternative route for carrying out the electric field of the propagating modes in the symmetric and antisymmetric constitutive parameters.

\subsection{Symmetric parameters case}

 For non-parallel vectors $\mathbf{d}$ and $\mathbf{n}$, the
electric field may be written as
\begin{eqnarray}
\mathbf{E} &=&\frac{E_{1}}{\left\vert \mathbf{d}\times \mathbf{n}\right\vert
}\mathbf{d}\times \mathbf{n}+\frac{E_{2}}{n\left\vert \mathbf{d}\times
	\mathbf{n}\right\vert }\mathbf{n\times }\left( \mathbf{d}\times \mathbf{n}%
\right)  \notag \\
&&-\frac{E_{1}}{\epsilon n^{2}}\tilde{\alpha}(\mathbf{n}\cdot \mathbf{d}%
)\left\vert \mathbf{d}\times \mathbf{n}\right\vert \mathbf{n,}
\label{symmetric_arb}
\end{eqnarray}
which does not supply, in general, 	transversal modes, that is,
$\mathbf{n}\cdot\mathbf{E}\neq 0$. However, for orthogonal vectors, $\mathbf{n}\cdot
\mathbf{d=0}$, Eq. (\ref{gauss2cc}) yields transversal
modes, $\mathbf{n}\cdot \mathbf{E}=0$, whose electric field can be written
as
	\begin{equation}
	\mathbf{E}={E_{1}}(\hat{\mathbf{d}}\times \hat{\mathbf{n}})+ {E_{2}}\hat{\mathbf{d}},\label{symmetric_orthogonaly}
	\end{equation}
	where $\hat{\mathbf{d}}={\mathbf{d}}/|{\mathbf{d}}|$ and $\hat{\mathbf{n}}={\mathbf{n}}/|{\mathbf{n}}|$. {In general, $E_{1}$ and $E_{2}$ are arbitrary constants, and for a normalized electric field, they satisfy $|E_{1}|^{2}+|E_{2}|^{2}=1$.}

For $\mathbf{n}$ and $\mathbf{d}$  nonparallel, the equation $\mathbb{M}\mathbf{E}=0$, with the electric field given in Eq. (\ref{symmetric_arb}), sets
	\begin{equation}
	\hspace{-0.1cm}{E}_{2}=E_{1}\frac{\mu \tilde{\alpha}(d^{2}\sin ^{2}\varphi )n_{\pm }}{n_{\pm }^{2}-\mu \epsilon }=E_{1}\mu \tilde{\alpha}(d^{2}\sin ^{2}\varphi)\Upsilon_{\pm } n_{\pm }, \label{elect_geral_1}
	\end{equation}
	where we have used Eqs.~(\ref{symmetric-mixed-case-2}) and (\ref{Upssilon}), and $n_\pm$ is given by Eq. (\ref{refractive-indices-symmetric-1b}). So the electric field [Eq.~(\ref{symmetric_arb})] {reads}
	\begin{eqnarray}
	\mathbf{E}_{\pm } &=&E_{0}\frac{\mathbf{d}\times \mathbf{n}_{\pm }}{%
		\left\vert \mathbf{d}\times \mathbf{n}_{\pm }\right\vert }+E_{0}\frac{\mu
		\tilde{\alpha}(d^{2}\sin ^{2}\varphi) \Upsilon _{\pm } }{\left\vert \mathbf{d}%
		\times \mathbf{n}_{\pm }\right\vert }\mathbf{n_{\pm }\times }\left( \mathbf{d%
	}\times \mathbf{n}_{\pm }\right)  \notag \\
	&&-E_{0}\frac{\tilde{\alpha}d^{2}\cos \varphi \sin \varphi }{\epsilon }%
	\mathbf{n}_\pm,  \label{symmetric_arbx}
	\end{eqnarray}
	setting $E_{1}=E_{0}$.  It provides the same conclusions obtained in Sec. \ref{anti_cc} for the general $\mathbf{d}$ configuration.

 For $\mathbf{n}$ and $\mathbf{d}$ orthogonal vectors, Eq.~(\ref{elect_geral_1}) simplifies as
	\begin{equation}
	E_{2}=E_{1}\frac{\mu \tilde{\alpha}d^2n_\pm }{ n^{2}_\pm-\mu \epsilon}=\pm \frac{\tilde{\alpha}}{|\tilde{\alpha}|}E_{1}.
	\end{equation}
	where $n_\pm$ is now given by Eq. (\ref{symmetric6T0}). So the electric field [Eq.~(\ref{symmetric_arbx})] becomes written as
	\begin{equation}
	\mathbf{E}_{\pm }=E_{0}\frac{\mathbf{d}\times \mathbf{n}_{\pm }}{|\mathbf{d}\times \mathbf{n}_{\pm }|}\pm
	E_{0}\frac{\tilde{\alpha}}{|\tilde{\alpha}|}\frac{\mathbf{d}}{d},
	\label{symmetric_orthogonalx}
	\end{equation}%
	providing the same conclusions obtained in the particular case of Sec. \ref{anti_bb}.

The vector formalism can also be used to alternatively express the permittivity tensor and the dispersion relation. In fact, replacing relations (\ref{symmetric1}) in the permittivity tensor (\ref{ex7}), {one writes}
\begin{equation}
\hat{\bar{\epsilon}}=\epsilon \mathbb{1}{-\tilde{\alpha}\mathbf{%
	d}\left( \mathbf{d}\times \mathbf{n}\right) ^{T}+\tilde{\beta}\left( \mathbf{	d}\times \mathbf{n}\right) \mathbf{d}^{T}},  \label{symmetric3b}
\end{equation}%
with $\hat{\bar{\epsilon}}=[{\bar{\epsilon}}_{ij}]$. In such
	a way the tensor $\mathbb{M}=[M_{ij}]$ [Eq.~\eqref{ex12}], takes the form
\begin{equation}
\mathbb{M}=(n^{2}-\mu \epsilon )\mathbb{1}-\mathbf{n}\mathbf{n}%
^{T}{+\mu \tilde{\alpha}\mathbf{d}(\mathbf{d}\times \mathbf{n})^{T}-\mu
\tilde{\beta}(\mathbf{d}\times \mathbf{n})\mathbf{d}^{T}.}  \label{symmetric4B}
\end{equation}
Evaluating $\mathrm{det}\mathbb{M}=0$, we read the dispersion relation (\ref{symmetric5}) in the form,
\begin{equation}
{\epsilon }\left( n^{2}-\mu \epsilon \right) ^{2}+\tilde{\alpha}\tilde{\beta}%
\mu \left[ \mu {\epsilon }d^{2}-(\mathbf{n}\cdot \mathbf{d})^{2}\right] \left\vert \mathbf{d}\times \mathbf{n}\right\vert ^{2}=0.
\label{symmetric5B}
\end{equation}

\subsection{Antisymmetric parameters case}

For nonparallel $\mathbf{a}$ and $\mathbf{n}$, the electric field is expressed as
\begin{equation}
\mathbf{\mathbf{E}}=\frac{E_{1}}{\left\vert \mathbf{a\times n}\right\vert }%
\mathbf{\left( \mathbf{a\times n}\right) -}\frac{\epsilon E_{3}}{n\left\vert
	\mathbf{a\times n}\right\vert ^{2}}\mathbf{\mathbf{n\times }\left( \mathbf{%
		a\times n}\right) }+\frac{E_{3}}{n}\mathbf{n}. \label{antisimetrico1}
\end{equation}%

The  equation $\mathbb{M}\mathbf{E}=0$ with the matrix \eqref{anti4} and the electric field given in Eq. (\ref{antisimetrico1}) provides the equations
	\begin{equation}
	E_{1}\left[ \frac{{}}{{}}n^{2}-\mu \epsilon +2\mu \mathbf{a}\cdot \mathbf{n}%
	\right] =0, \label{rant_1}
	\end{equation}
	\begin{equation}
	E_{3}\left[ \epsilon\left( n^{2}-\mu \epsilon +2\mu \mathbf{a}\cdot \mathbf{n}%
	\right) + \mu\left\vert \mathbf{a\times n}\right\vert ^{2}\right] = 0. \label{rant_2}
	\end{equation}
	where the expressions in the brackets are the dispersion relations obtained in Eq. (\ref{anti5b}) when $\mathbf{b}=\mathbf{a}$.

	If the dispersion relation in Eq. (\ref{rant_1}), $n^{2}-\mu \epsilon +2\mu \mathbf{a}\cdot \mathbf{n}=0$, is valid, then $E_1$ remains arbitrary and $E_3=0$. Thus, the electric field mode (\ref{antisimetrico1}) becomes transversal to the propagation direction, that is,
	\begin{equation}
	\mathbf{\mathbf{E}}=\frac{E_{1}}{\left\vert \mathbf{a\times n}\right\vert }%
	( \mathbf{a}\times \mathbf{n}),
	\end{equation}
	being in accordance with the result (\ref{anti18A}) of the case analyzed in Sec. \ref{case1ants}. On the other hand, if the dispersion relation in Eq. (\ref{rant_2}) holds, $\epsilon\left( n^{2}-\mu \epsilon +2\mu \mathbf{a}\cdot \mathbf{n}%
	\right) + \mu\left\vert \mathbf{a\times n}\right\vert ^{2}=0$, then $E_1=0$ and $E_3 $ remains arbitrary, so that the electric field modes (\ref{antisimetrico1}) are no longer transverse,
	\begin{equation}
	\mathbf{\mathbf{E}}=\mathbf{-}\frac{\epsilon E_{3}}{n\left\vert \mathbf{%
			a\times n}\right\vert ^{2}}\mathbf{\mathbf{n\times }\left( \mathbf{a\times n}%
		\right) }+\frac{E_{3}}{n}\mathbf{n}.
	\end{equation}
	It is in accordance with the result (\ref{anti18B}) of the case analyzed in Sec. \ref{case1ants}.

Replacing Eqs. \eqref{anti1} and \eqref{anti2} in Eq.~\eqref{ex7}, {
one obtains} the following extended electric permittivity tensor:
\begin{equation}
\bar{\epsilon}=\left( \epsilon -\mathbf{c}\cdot \mathbf{n}\right) \mathbb{1}+
\mathbf{b}\mathbf{n}^{T}+\mathbf{n}\mathbf{a}^{T},  \label{anti3b}
\end{equation}%
where $\mathbf{c}=\mathbf{a}+\mathbf{b}$
is a real vector [see Eq. (\ref{a=b*})]. All the pieces of
	the new electric permittivity coming from antisymmetric parameters contain
direction-dependent terms: {$(\mathbf{c}\cdot \mathbf{n})$,  $b_{i}n_{j}$, $n_{i}a_{j}$.} Now, the explicit form of the matrix $\mathbb{M}=[M_{ij}]$ is
\begin{equation}
\mathbb{M}=[n^{2}-\mu \epsilon +\mu(\mathbf{c}\cdot \mathbf{n})] \mathbb{1} -\mathbf{n}\mathbf{n}^{T}-\mu \mathbf{b}\mathbf{n}^{T}-\mu
\mathbf{n}\mathbf{a}^{T}. \label{anti4b}
\end{equation}%

\section{\label{AppendixB} Rotatory power and dichroism coefficient}

Relation (\ref{eq:rotatory-power1A}) holds for the situation that the
propagating modes are given by circularly polarized waves (LCP and RCP). To
examine the effect of the optical activity on the modes, we start from a linearly polarized wave propagating through a medium along the $z$ axis. As is
well known, a wave with linear polarization,
\begin{subequations}
\begin{equation}
\mathbf{E}_{i}= \mathbf{E}_{0i} \mathrm{e}^{\mathrm{i} (kz-\omega t)}\,,
\label{rotation-1}
\end{equation}
can be split into two circularly polarized waves,
\begin{align}
\mathbf{E}_{0i}=%
\begin{pmatrix}
1 \\
0 \\
0%
\end{pmatrix}
= \frac{1}{2}
\begin{pmatrix}
1 \\
-\mathrm{i} \\
0%
\end{pmatrix}
+ \frac{1}{2}
\begin{pmatrix}
1 \\
\mathrm{i} \\
0%
\end{pmatrix}%
\,,  \label{rotation-2}
\end{align}
corresponding to the sum of RCP and LCP waves, respectively. After the
initial wave passes through a distance $z$ in the medium, the final electric
field can be obtained as the combination of two components, $\mathbf{E}_{+}$
and $\mathbf{E}_{-}$, with the wave vectors $\mathbf{k}_{+}$ and $\mathbf{k}%
_{-}$, respectively. One then has
\end{subequations}
\begin{align}
\mathbf{E}_{f} &= \mathbf{E}_{+} \mathrm{e}^{\mathrm{i}( k_{+} z - \omega
t)}+ \mathbf{E}_{-} \mathrm{e}^{\mathrm{i} (k_{-} z - \omega t)}  \notag \\
&=\frac{1}{2}
\begin{pmatrix}
1 \\
\mathrm{i} \\
0%
\end{pmatrix}
\mathrm{e}^{\mathrm{i} k_{+} z} \mathrm{e}^{-\mathrm{i}\omega t} + \frac{1}{2%
}
\begin{pmatrix}
1 \\
-\mathrm{i} \\
0%
\end{pmatrix}
\mathrm{e}^{\mathrm{i}k_{-} z} \mathrm{e}^{-\mathrm{i}\omega t}\,,
\label{rotation-3}
\end{align}
which can be cast into the form 
%
%
\begin{subequations}
\begin{align}
\mathbf{E}_{f} &= \frac{1}{2} \mathrm{e}^{\mathrm{i} \psi} \mathrm{e}^{-%
\mathrm{i}\omega t} \left[ \mathrm{e}^{-\mathrm{i} \theta}
\begin{pmatrix}
1 \\
\mathrm{i} \\
0%
\end{pmatrix}
+ \mathrm{e}^{\mathrm{i} \theta}
\begin{pmatrix}
1 \\
-\mathrm{i} \\
0%
\end{pmatrix}%
\right]  \notag \\
&=\mathrm{e}^{\mathrm{i}\psi} \mathrm{e}^{-\mathrm{i}\omega t}
\begin{pmatrix}
\cos \theta \\
\sin \theta \\
0%
\end{pmatrix}%
\,,  \label{rotation-10}
\end{align}
with the quantities
\begin{align}
\theta &= -\frac{( k_{+}-k_{-})z}{2}\,, \displaybreak[0]  \label{rotation-5}
\\
\psi &= \frac{(k_{+}+k_{-})z}{2}\,.  \label{rotation-6}
\end{align}
Notice that Eq. \eqref{rotation-10} describes a linearly polarized wave whose
polarization vector is rotated by an angle $\theta$. From Eq.~\eqref{rotation-5}%
, one obtains
\end{subequations}
\begin{equation}
\theta=- \frac{ (n_{+}-n_{-}) z \omega}{2}\,,  \label{rotation-11}
\end{equation}
where we have used $\mathbf{k}=\omega \mathbf{n}$. In general, the
refractive indices can be complex quantities. Because of this, one can infer
from Eq. \eqref{rotation-11}
\begin{equation}
\frac{\theta}{z}= -\frac{\omega}{2} \left[\mathrm{Re} (n_{+}) + \mathrm{i}
\mathrm{Im}(n_{+}) - \mathrm{Re} (n_{-}) - \mathrm{i} \mathrm{Im}(n_{-})%
\right]\,,  \label{rotation-12}
\end{equation}
from which we define the specific rotatory power,
\begin{equation}
\delta = - \frac{[ \mathrm{Re}(n_{+})-\mathrm{Re}(n_{-}) ] \omega}{2}\,,
\label{eq:rotatory-power12}
\end{equation}
as well as the dichroism coefficient,
\begin{equation}
\delta_{\mathrm{d}}= - \frac{ [\mathrm{Im}(n_{+})- \mathrm{Im}(n_{-}) ]
\omega}{2}\,.  \label{rotation-12b}
\end{equation}
Notice that when the medium is nonbirefringent, $\theta=0$ and $\psi=k z$.
Then, the form \eqref{rotation-1} is recovered from Eq.\eqref{rotation-10}.

\section{\label{Appendix-parameters-relations} Relations for constitutive
parameters}

The energy conservation in electromagnetic systems is {established by the
Poynting theorem, presented in Eq.~\eqref{eq:poynting-theorem-3} as}
\begin{equation}
\mathrm{Re}\left[ \mathrm{i}\omega \left( \mathbf{E}\cdot \mathbf{D}^{\ast }-%
\mathbf{H}^{\ast }\cdot \mathbf{B}\right) \right] =0,  \label{poynting-30}
\end{equation}%
or
\begin{equation}
\mathrm{Re}\left[ \mathrm{i}\omega \left( \mathbf{D}^{\dag }\mathbf{E}-%
\mathbf{{H^{\dag }B}}\right) \right] =0.  \label{poynting-3}
\end{equation}
We can now consider a medium described by the following constitutive
relations

\begin{subequations}
\label{poynting-21}
\begin{align}
\mathbf{D}& =\hat{\epsilon} \mathbf{E} +\hat\lambda \mathbf{H} ,
\label{poynting-21-a} \\
\mathbf{B} & =\hat\mu \mathbf{H} +\hat\gamma \mathbf{E} .
\label{poynting-21-b}
\end{align}
\end{subequations}
with $\hat{\tilde\epsilon}$, $\hat{\lambda}$, $\hat{\mu}$ and $\hat{\gamma}$
being nonsingular $3\times 3$ complex matrices. Following the same previous procedure, Eq.~\eqref{poynting-3} yields
\begin{align}
\mathbf{E}^\dag\left( \hat{\epsilon}^\dag-\hat{\epsilon}%
\right) \mathbf{E} +\mathbf{H}^\dag \left(\hat\mu^\dag -\hat\mu\right)
\mathbf{H}& +  \notag \\
+\mathbf{H}^\dag \left(\hat\lambda^\dag -\hat\gamma\right) \mathbf{E} -%
\mathbf{E}^\dag \left(\hat\lambda -\hat\gamma^\dag\right) \mathbf{H}& =0.
\label{poynting-22}
\end{align}
In this case, for arbitrary fields, in order to {be consistent with} Eq.~\eqref{poynting-22}, one finds
\begin{equation}
\hat{\epsilon}^\dag=\hat{\epsilon}, \quad \hat\mu^\dag =\hat\mu, \quad \hat\lambda =\hat\gamma^\dag,  \label{poynting-23}
\end{equation}
conditions compatible with energy conservation.

\end{document}